\documentstyle[prl,aps,floats,epsfig]{revtex}
\newcommand{\be}{\begin{equation}}
\newcommand{\ee}{\end{equation}}
\newcommand{\bees}{\begin{eqnarray}}
\newcommand{\ees}{\end{eqnarray}}
\newcommand{\ra}{\rightarrow}
\def\lsim{\raise 0.4ex\hbox{$<$}\kern -0.8em\lower 0.62
ex\hbox{$\sim$}}
\def\gsim{\raise 0.4ex\hbox{$>$}\kern -0.7em\lower 0.62
ex\hbox{$\sim$}}
\newcommand{\mpl}{M_{\rm Pl}}
\newcommand{\mgut}{M_{\rm GUT}}

\newcommand{\tpl}{t_{\rm Pl}}

\newcommand{\ogw}{\Omega_{\rm gw}}
\newcommand{\hogw}{h_0^2\Omega_{\rm gw}}
\newcommand{\hc}{h_c(f)}

\begin{document}

\par
\begingroup
%\twocolumn[%

\begin{flushright}
VIR-NOT-PIS-1390-113\\
 IFUP-TH 58/97\\
 March 1998\\
%  gr-qc/9803028
\end{flushright}

\vspace{5mm}

{\large\bf\centering\ignorespaces
High-Energy Physics with  Gravitational-Wave Experiments
\vskip2.5pt}
{\dimen0=-\prevdepth \advance\dimen0 by23pt
\nointerlineskip \rm\centering
\vrule height\dimen0 width0pt\relax\ignorespaces
Michele Maggiore
\par}
{\small\it\centering\ignorespaces

\vspace*{3mm}

INFN, sezione di Pisa, and
 Dipartimento di Fisica, Universit\`{a} di Pisa,\\
piazza Torricelli 2, I-56100 Pisa, Italy\par}

\par
\bgroup
\leftskip=0.10753\textwidth \rightskip\leftskip
\dimen0=-\prevdepth \advance\dimen0 by17.5pt \nointerlineskip
\small\vrule width 0pt height\dimen0 \relax

We discuss  the possible relevance
of gravitational-wave (GW) experiments for physics at very high
energy. We examine whether, from the experience gained with the
computations of various specific relic GW backgrounds, we can extract 
statements and order of magnitude estimates that are as much 
as possible model-independent, and we try to distinguish between
general conclusions and results related to specific  cosmological
mechanisms.  We examine the statement
that the  Virgo/LIGO experiments
probe the Universe at temperatures $T\sim
10^{7}-10^{10}$ GeV (or timescales $t\sim 10^{-20}-10^{-26}$ sec) and 
we consider the possibility that they could actually
probe the Universe at much higher energy scales,
including the typical scales of grand unification, string
theory and quantum gravity. We consider possible scenarios,
depending  on how the inflationary paradigm is implemented.
We discuss the prospects for detection 
with present and planned experiments. In particular, a
second Virgo interferometer correlated with the planned one, and
located within a few tens of kilometers from the first, 
could reach an interesting sensitivity for
stochastic GWs of cosmological origin.

\par\egroup

\thispagestyle{plain}
\endgroup

\section{Introduction}
The energy range between the grand unification scale $M_{\rm GUT}\sim
10^{16}$ GeV and the Planck scale $\mpl\simeq 1.22\times 10^{19}$ GeV
is crucial for fundamental physical questions and for testing current
ideas about grand unification, quantum gravity, string theory.
Experimental results  in this  energy range are  of course
very difficult to obtain. From the particle physics point of view,
there are basically two important experimental
 results that can be translated into statements about this
energy range (see e.g.~\cite{rev} for
recent reviews): (i) the accurate measurement of gauge
coupling constants at LEP that, combined with their running 
with energy, shows the unification of the couplings at the
scale $\mgut$, provided that the running is computed including the
supersymmetric particles in the low energy spectrum. 
And (ii) the negative results on proton
decay; the lower limit on the inverse of the partial decay width for
the processes $p\ra e^+\pi^0$ and $p\ra K^+\bar{\nu}$ are 
$5.5\times 10^{32}$ yr and $1.0\times 10^{32}$ yr, respectively, and
imply a lower bound  $\mgut \gsim 10^{15}$ GeV, which excludes
non-supersymmetric SU(5) unification. Further improvement is expected
from the SuperKamiokande experiment, which should probe lifetimes 
$\sim 10^{34}$ yr.

From the cosmological point of view, informations on this energy range
can only come from particles which decoupled  from the primordial
plasma at  very early time.
Particles which stay in thermal equilibrium down to a
decoupling temperature $T_{\rm dec}$ can only carry informations on
the state of the Universe at
 $E\sim T_{\rm dec}$. All informations on physics at higher
energies has  in fact been obliterated by the successive interactions. 

The condition for thermal equilibrium is that the rate $\Gamma$
of the processes that mantain equilibrium be larger than the 
rate of expansion of the Universe, as measured by the Hubble
parameter $H$~\cite{KT}. 
The rate is given by $\Gamma =n\sigma |v|$
where $n$ is the number density of the particle in question,
and for massless or light particles in equilibrium at a temperature $T$,
$n\sim T^3$;  $|v|\sim 1$ is the typical velocity and
$\sigma$ is the cross-section of the process. 
Consider for instance the weakly interacting neutrinos. In this
case the equilibrium is mantained, e.g., by electron-neutrino
scattering, and at energies below the $W$ mass
 $\sigma\sim G_F^2\langle E^2\rangle
\sim G_F^2T^2$ where $G_F$ is the
 Fermi constant and $\langle E^2\rangle$ is the average energy squared.
 The Hubble parameter during the radiation
dominated era is related to the temperature by $H\sim
 T^2/\mpl$. Therefore~\cite{KT}
\be
\left(\frac{\Gamma}{H}\right)_{\rm neutrino}
\sim \frac{G_F^2T^5}{T^2/\mpl}\simeq\left(
\frac{T}{1\rm MeV}\right)^3\, .
\ee
Even the weakly interacting neutrinos, therefore, cannot carry
informations on the state of the Universe at temperatures larger than
approximately 1 MeV.
If we repeat the above computation for gravitons, the Fermi constant
$G_F$ is replaced by Newton constant $G=1/\mpl^2$ (we always use
units $\hbar =c=k_B=1$) and at energies below the Planck mass
\be
\left(\frac{\Gamma}{H}\right)_{\rm graviton}
\sim \left(\frac{T}{\mpl}\right)^3\, .
\ee
The gravitons are therefore decoupled below the Planck scale. (At
the Planck scale the above estimate of the cross section
is not valid and nothing can be said without a quantum theory of
gravity).
It follows that relic gravitational waves  are a potential source of
informations on very high-energy physics. Gravitational waves
produced in the very early Universe have not lost memory of 
the conditions in which they have been produced, as it happened to
all other particles, but still retain in their spectrum, typical
frequency and intensity, important informations on the state of the
very early Universe,  and therefore on 
physics at correspondingly high energies, 
which cannot be accessed experimentally in any
other way.
It is also clear that the property of gravitational waves that
makes them so interesting, 
i.e. their extremely small cross section, is also
responsible for the difficulties of the experimental
detection.\footnote{Thinking in terms of cross-sections, one is lead
to ask  how comes that
 gravitons could be detectable altogheter, since the graviton-matter
cross section
is smaller than the neutrino-matter cross section,
at energies below the $W$-mass, by a
factor $G^2/G_F^2\sim 10^{-67}$ and 
 neutrinos are
already so difficult to detect. The answer is that gravitons are
bosons, and therefore their occupation number per cell of phase space 
can be
$n_k\gg 1$; we will see below that in interesting cases, in the relic
stochastic background we can have
$n_k\sim 10^{40}$ or larger, and the squared amplitude 
for exciting a given mode ot the detector grows as $n_k^2$.
So, we will never really detect gravitons, but
rather classical gravitational waves. 
Neutrinos, in contrast, are fermions and for them $n_k\leq 1$.}

 With the very limited
experimental informations that we have on the very high energy
region,  $\mgut \lsim E\lsim \mpl$,
it is unlikely that  theorists will be able to
foresee all the interesting sources of relic  stochastic background,
let alone to compute their spectra. This is particularly
clear in the Planckian or
string theory domain where, even if we succeed in predicting some
interesting physical effects, in general we cannot
 compute them reliably. So, despite the large efforts that have
been devoted to understanding possible sources, it is still quite
possible that, if a relic background of gravitational waves will be
detected, its physical origin will be a surprise.
In this case a model-independent analysis of what we can
expect might be useful.

In this paper we discuss  whether, from the
experience gained with various specific computations of relic
backgrounds,  it is possible to extract statements
or order of magnitude estimates
which are as much as possible model-independent. 
These estimates would constitute a sort of minimal set of naive
expectations, that could give some orientation, independently of the
uncertainties and intricacies of the specific cosmological models.
We discuss  typical values of
the frequencies involved and of the  expected intensity of
the background gravitational radiation, and we try to distinguish 
between statements that 
are relatively model-independent and results  specific to given
models. 

The paper is written having in mind a reader interested in
gravitational-wave detection
but not necessarily competent in early Universe
cosmology nor in physics at the string or Planck scale, and a number
of more technical remarks are relegated in footnotes and in 
an appendix. We have also
tried to be self-contained and we have attempted to summarize and
occasionally clean up many formulas and numerical estimates appearing
in the literature. The organization of the paper
is as follows. In sect.~2 we introduce the variables
most commonly used to describe a stochastic background of
gravitational waves. We give a detailed derivation of the relation
between exact formulas for the signal-to-noise ratio, and approximate
but simpler characterizations of the characteristic
amplitude and of the noise. The former variables are convenient in
theoretical computations while the latter are commonly used by
experimentalists, so it is worthwhile to understand in some details
their relations.
In sect.~3 we apply these formulas to compute the
 sensitivity to a stochastic background that could be obtained
with a second Virgo interferometer correlated with the first,
and we compare with various others detectors. We find that in 
the Virgo-Virgo  case
the  noise which would give the dominant limitation to
the measurement of a stochastic
background is the mirror thermal noise, and we give the sensitivity
for different forms of the relic GW spectrum.
In sect.~4 and 5 we discuss estimates of the typical
frequency scales. We examine the statements leading to the conclusion
that Virgo/LIGO will explore the Universe at temperatures $T\sim 10^7$
GeV (sect.~4), and in the appendix
we discuss some qualifications to this statement.
In sect.~5  we discuss the possibility to reach much higher energy
scales, including the typical scales of grand unification and quantum
gravity. In sect.~6 we discuss different scenarios, depending on how
the inflationary paradigm is implemented. 
Characteristic values of the intensity
of the spectrum and existing limits and predictions are discussed in
sect.~7. Sect.~8 contains the conclusions.

\section{Definitions}
\subsection{$\ogw (f)$ and the optimal SNR}
The intensity of  a stochastic background of gravitational waves (GWs)
can be characterized by the dimensionless quantity
\be
\ogw (f)=\frac{1}{\rho_c}\,\frac{d\rho_{\rm gw}}{d\log f}\, ,
\ee
where $\rho_{\rm gw}$ is the energy density of the stochastic
background of gravitational waves, $f$ is the frequency ($\omega =2\pi
f$) and $\rho_c$ is the present value of the
critical energy density for closing the Universe. In terms of the present
value of the Hubble constant $H_0$, the critical density is given by
\be\label{rhoc}
\rho_c =\frac{3H_0^2}{8\pi G}\, .
\ee
The  value of $H_0$ is usually written as $H_0=h_0\times 100 $
km/(sec--Mpc), where $h_0$ parametrizes the existing experimental
uncertainty.  Ref.~\cite{PDG} gives a value
 $0.5<h_0<0.85$. In the last
few years there has been a constant trend toward lower values of
$h_0$ and typical estimates are now in the range $0.55<h_0<0.60$
or, more conservatively, $0.50<h_0<0.65$. For instance
ref.~\cite{San}, using the method of type IA supernovae,
 gives two slightly different estimates
$h_0=0.56\pm 0.04$ and $h_0=0.58\pm 0.04$.
Ref.~\cite{Tri}, with the same method,
finds $h_0=0.60\pm 0.05$ and ref.~\cite{KK}, using a gravitational
lens, finds $h_0=0.51\pm 0.14$. 
The spread of values obtained gives an idea of
the systematic errors involved.

It is not very convenient to normalize $\rho_{\rm gw}$
to a quantity, $\rho_c$,
which is uncertain: this uncertainty would appear
in all the subsequent formulas, although it has nothing to do with the 
uncertainties on the GW background. Therefore, we
rather characterize the stochastic GW background with
the quantity $\hogw (f)$, which is independent of $h_0$. All theoretical
computations of a relic GW spectrum are actually  computations of 
$d\rho_{\rm gw}/d\log f$ and  are  independent of the
uncertainty on $H_0$. Therefore the result of these computations is
 expressed in terms of $\hogw$, rather than of
$\ogw$.\footnote{This simple point has occasionally been missed in the
literature, where one can find the statement that, for small values of
$H_0$, $\ogw$ is larger and therefore easier to detect. Of course, it
is larger only because it has been normalized using a smaller
quantity.} 

To detect a stochastic GW background the optimal strategy consists in
performing a correlation between two (or more) detectors, since, as we
will discuss below, the
signal  will be far too low to exceed the noise
level in any existing or planned 
single detector (with the exception of the space interferometer LISA,
see below).
The strategy has been discussed
in refs.~\cite{Mic,Chr,Fla,VCCO}, and a 
clear review is ref.~\cite{All}. Let us recall
the main points of the analysis.
The output of any single detector is of the form
$s_i(t)=h_i(t)+n_i(t)$, where $i=1,2$ labels the detector, and the
output $s_i(t)$ is made up of a noise $n_i(t)$ and possibly a signal
$h_i(t)$. In the typical situation, $h_i\ll n_i$.
We can correlate the two outputs defining
\be\label{S}
S=\int_{-T/2}^{T/2}dt\int_{-T/2}^{T/2}dt'\,
s_1(t)s_2(t')Q(t-t')\, ,
\ee
where $T$ is the total integration time (e.g. one year) and $Q$ a
filter function. If the noises in the two detectors are uncorrelated,
the ensemble average of the Fourier components of the noise satisfies
\be\label{Sn}
\langle \tilde{n}_i^*(f)\tilde{n}_j(f')\rangle =
\delta(f-f')\delta_{ij}\frac{1}{2}S_n^{(i)}(|f|)\, .
\ee
The above equation defines the
functions $S_n^{(i)}(|f|)$, with dimensions Hz$^{-1}$. 
The factor $1/2$ is conventionally inserted in the definition
so that the total noise power is obtained integrating $S_n(f)$ over the
physical range $0\leq f<\infty$, rather than from $-\infty$ to
$\infty$. 
The noise level of the  detector labelled by $i$ is therefore measured by
$\tilde{h}_f^{(i)}\equiv\sqrt{S_n^{(i)}}$, with dimensions
Hz$^{-1/2}$. The function $S_n$ is known as the square spectral noise
density.\footnote{Unfortunately there is not much agreement
about notations in the literature. The spectral noise density,
that we denote by $S_n(f)$ following e.g. ref.~\cite{Fla},
is called $P(f)$ in ref.~\cite{All}. Other authors
use  the notation $S_h(f)$, which we instead reserve 
for the spectral density of the signal. To make things worse, $S_n$ is
sometime defined with or without the factor 1/2 in eq.~(\ref{Sn}).}

As discussed in
refs.~\cite{Mic,Chr,Fla,VCCO,All}, in the limit $h_i\ll n_i$,
for any given form of the signal,
i.e. for any given functional form of $\hogw (f)$, it is possible to find
explicitly the filter function $Q(t)$ which maximizes the
signal-to-noise ratio (SNR). In the case
of L-shaped interferometers the corresponding value of the optimal
SNR  turns out to be (see e.g. ref.~\cite{All}, eq.(43))
\be\label{SNR}
{\rm SNR}=\left[ \left(\frac{9H_0^4}{50\pi^4}\right)\, 
T\int_0^{\infty}df\,
\frac{\gamma^2(f)\ogw^2(f)}{f^6S_n^{(1)}(f)S_n^{(2)}(f)}
\right]^{1/4}\, .
\ee
We have
taken into account the fact that what has been called $S$ in
eq.~(\ref{S}) is quadratic in the signals and, with usual definitions,
it contributes to the SNR squared. This differs from the convention
used in ref.~\cite{All}.
The function $\gamma (f)$ is called the overlap function. It takes
into account the difference in location and orientation of the two
detectors. It has been computed for the various pairs of LIGO1, LIGO2,
Virgo and GEO detectors~\cite{Fla}. For detectors very close and parallel,
$\gamma (f)=1$. Basically, $\gamma (f)$
  cuts off the integrand in eq.~(\ref{SNR}) 
at a frequency $2\pi f$ of the order of the inverse separation between
the two detectors. For the two LIGO detectors, this cutoff is around
60 Hz. We will discuss $\gamma (f)$ 
in sect.~2C, where we will also comment on
the modifications needed for different geometries.

In principle the expression for the SNR, eq.~(\ref{SNR}), is all
that we need in order to discuss the possibility of detection of a
given GW background. However it is useful, for order of magnitude
estimates and for intuitive understanding, to express the SNR  in
terms of a characteristic amplitude of the stochastic GW background
and of a characteristic noise level, although, as we will see, the
latter is a quantity that describes the noise only approximately, in
contrast to eq.~(\ref{SNR}) which is exact.
We will introduce these quantities in the next two subsections.

\subsection{The characteristic amplitude}

A stochastic GW at a given point $\vec{x}=0$ can be expanded, in the
transverse traceless gauge, as (we follow  the notations of 
ref.~\cite{All}, app.A)
\be\label{hab}
h_{ab}(t)=\sum_{A=+,\times}\int_{-\infty}^{\infty}df\int d\hat{\Omega}
\,\tilde{h}_A(f,\hat{\Omega})\exp (2\pi ift)e_{ab}^A(\hat{\Omega})\, ,
\ee
where $\tilde{h}_A(-f,\hat{\Omega})=\tilde{h}_A^*(f,\hat{\Omega})$. 
$\hat{\Omega}$ is a unit vector representing the direction of
propagation of the wave and $d\hat{\Omega}=d\cos\theta d\phi$.
The polarization tensors can be written as
$e_{ab}^+(\hat{\Omega})=m_am_b-n_an_b$ and
$e_{ab}^{\times}(\hat{\Omega})=m_an_b+n_am_b$,
with $\hat{m},\hat{n}$ unit vectors ortogonal to 
$\hat{\Omega}$ and to each other. With these definitions,
$e^A_{ab}(\hat{\Omega})e^{A',ab}(\hat{\Omega})=2\delta^{AA'}$. For a
stochastic background,  assumed to be isotropic, unpolarized and
stationary (see~\cite{All,AR} for a discussion of these assumptions)
the ensemble average of the Fourier amplitudes
can be written as
\be\label{ave}
\langle \tilde{h}_A^*(f,\hat{\Omega})
\tilde{h}_{A'}(f',\hat{\Omega}')\rangle =
\delta (f-f')\frac{1}{4\pi}
\delta^2(\hat{\Omega},\hat{\Omega}')\delta_{AA'}
\frac{1}{2}S_h(f)\, ,
\ee
where $\delta^2(\hat{\Omega},\hat{\Omega}')=\delta (\phi -\phi ')
\delta (\cos\theta -\cos\theta ')$. The function $S_h(f)$ defined by
the above equation
 has dimensions Hz$^{-1}$ and  satisfies $S_h(f)=S_h(-f)$. 
The factor 1/2 is conventionally inserted  in the definition of
$S_h$ in order to compensate for the fact
that the integration variable $f$
in eq.~(\ref{hab}) ranges between $-\infty$ and
$+\infty$ rather than over the physical domain $0\leq f<\infty$.
The factor $1/(4\pi )$ is inserted so that, integrating the 
left-hand side over $d\hat{\Omega}$ and over $d\hat{\Omega} '$, we get 
$\delta (f-f')\delta_{AA'}(1/2)S_h(f)$. With this normalization, 
$S_h(f)$ is therefore the quantity to be compared with the noise level
$S_n(f)$ defined in eq.~(\ref{Sn}). 
Using eqs.~(\ref{hab},\ref{ave}) we get
\be\label{hc1}
\langle h_{ab}(t)h^{ab}(t)\rangle =
2\int_{-\infty}^{\infty}df\, S_h(f)=
4\int_{f=0}^{f=\infty}d(\log f)\,\, fS_h(f)\, .
\ee
We  now define the characteristic amplitude $h_c(f)$ from
\be\label{hc2}
\langle h_{ab}(t)h^{ab}(t)\rangle =
2 \int_{f=0}^{f=\infty}d(\log f)\,\, h_c^2(f)\, .
\ee
Note that $h_c(f)$ is dimensionless, and represents a characteristic
value of the amplitude, per unit logarithmic interval of frequency.
The factor of two on the right-hand side of eq.~(\ref{hc2}) 
is inserted for the following reason. 
The response of the detector to a single wave with amplitudes
$h_+,h_{\times}$ is of the form (Ref.~\cite{Th}, eqs.~(26) and (103))
$h(t)=F_+h_++F_{\times}h_{\times}$.
For an interferometer,  $h(t)=\Delta l(t)/L$ where $l$ is the
difference in arm lengths.  The function $h(t)$ is also called the
gravitational wave strain acting on the detector.
The functions  $F_{+,\times}$ are known as detector
pattern functions, and $0\leq |F_{+,\times}|\leq 1$. 
They depend  on three angles, that determine
the direction of arrival of the GW and  its polarization.
For any quadrupole-beam-pattern GW
detector  we have~(ref.\cite{Th}, pg. 369)
$\langle F_+^2\rangle_{\hat{\Omega}}
 =\langle F_{\times}^2\rangle_{\hat{\Omega}}$ and
$\langle F_+F_{\times}\rangle_{\hat{\Omega}} =0$, 
where
\be
\langle\ldots\rangle_{\hat{\Omega}}=
\int \frac{d\hat{\Omega}}{4\pi}\, \frac{d\psi}{2\pi} 
(\ldots )
\ee
denotes an average 
over the direction of propagation of the wave, $d\hat{\Omega}=
d\cos\theta d\phi$, and
over the angle $\psi$ that gives the preferred frame $(x',y')$
where the wave in
the transverse traceless gauge takes the simple form
$h_{x'x'}=-h_{y'y'}$, $h_{x'y'}=h_{y'x'}$ (see~\cite{Th}, pg.~367).
This average should
not to be confused with $\langle\ldots\rangle$ which is the time
average, i.e., in Fourier space,
the ensemble average of eq.~(\ref{ave}). Therefore, 
$\langle h^2(t)\rangle_{\hat{\Omega}}=
\langle F_+^2\rangle_{\hat{\Omega}}h_+^2+
\langle F_{\times}^2\rangle_{\hat{\Omega}}h_{\times}^2$.
In a unpolarized stochastic background 
$\langle h_+^2\rangle =\langle h_{\times}^2\rangle$
(where $\langle\ldots\rangle$ is the ensemble average of
eq.~(\ref{ave})) and therefore, since
$\langle F_+^2\rangle_{\hat{\Omega}}=
\langle F_{\times}^2\rangle_{\hat{\Omega}}$, 
the two terms in $ h^2(t)$, after averaging over the angles and
over time,
 give the same contribution. This motivates the factor of
two in the definition of $h_c^2(f)$, eq.~(\ref{hc2}).  
For the same reason, we could  insert a
factor $\langle F_+^2\rangle_{\hat{\Omega}}$ in the definition of
$h_c^2(f)$. However, what is really meaningful  is not a characterization
of the signal $h_c(f)$ nor of the noise level $h_n(f)$ separately,
but rather the signal-to-noise ratio discussed in the previous
subsection.  We are trying to
express the SNR in terms of a ratio of two quantities, $\hc /h_n(f)$.
So, we are free to  arbitrarily move factors from $h_c$ to $h_n$ as long
as $h_c/h_n$ is unchanged. However, it is convenient to collect in
$h_c$ all factors related to the source, and in $h_n$ all
factors  related to the detectors. We do not
insert a factor $\langle F_+^2\rangle_{\hat{\Omega}}^{1/2}$ in the
definition of $h_c$, and therefore, automatically, we
will obtain a factor
$\langle F_+^2\rangle_{\hat{\Omega}}^{-1/2}$ in $h_n$,
see  eqs.~(\ref{hn},\ref{gam2}) below. The 
same is true for numerical factors like the factor of two just
discussed. We could as well have neglected 
it in $h_c^2$ and we would find
an additional factor $1/\sqrt{2}$ in $h_n$.
We will discuss $h_n$  in the next subsection.

Comparing eqs.~(\ref{hc1}) and
(\ref{hc2}), we get
\be\label{hc3}
h_c^2(f)= 2fS_h(f)\, .
\ee
We now wish to relate $\hc$ and $\hogw (f)$. The starting point is the
expression for the energy density of gravitational waves,
given by the $00$-component of the energy-momentum tensor. 
The energy-momentum tensor of a GW cannot be localized inside a
single wavelength
(see e.g.~ref.\cite{MTW}, sects.~20.4 and 35.7 for a careful
discussion) but it can be defined  with a spatial averaging over
several wavelenghts:
\be\label{rho1}
\rho_{\rm gw}=\frac{1}{32\pi G}\langle \dot{h}_{ab}\dot{h}^{ab}
\rangle\, .
\ee
For a stochastic background, the spatial average over a few wavelengths
is the same as a time average at a given point, 
which, in Fourier space, is the 
ensemble average performed using eq.~(\ref{ave}). We therefore
insert eq.~(\ref{hab}) into eq.~(\ref{rho1}) and use
eq.~(\ref{ave}). The result is 
\be\label{rho2}
\rho_{\rm gw}=\frac{4}{32\pi G}
\int_{f=0}^{f=\infty}d(\log f)\,\, f (2\pi f)^2S_h(f)\, ,
\ee
so that 
\be\label{rho3}
\frac{d\rho_{\rm gw}}{d\log f}=\frac{\pi}{2G}f^3S_h(f)\, .
\ee
Comparing eqs.~(\ref{rho3}) and (\ref{hc3}) we get the important
relation
\be\label{rho4}
\frac{d\rho_{\rm gw}}{d\log f}=\frac{\pi}{4G}f^2h_c^2(f)\, ,
\ee
or, dividing by the critical density $\rho_c$,
\be\label{ogwhc}
\ogw (f)=\frac{2\pi^2}{3H_0^2} f^2h_c^2(f)\, .
\ee
Inserting the numerical
value of $H_0$, we find (ref.~\cite{Th}, eq.~(65))
\be\label{rho5}
h_c(f)\simeq 1.263\times 10^{-18}\,
\left(\frac{{\rm 1 Hz}}{f}\right)\sqrt{\hogw (f)}\,\, .
\ee
Using eqs.~(\ref{hc3},\ref{ogwhc}) we can also write
$\ogw (f)=(4\pi^2/3H_0^2) f^3S_h(f)$. Using this relation,
and defining $S_n(f)=(S_n^{(1)}(f)S_n^{(2)}(f))^{1/2}$,
eq.~(\ref{SNR}) can be written in a  more transparent form,
\be\label{SNR2}
{\rm SNR}=\left[ 2 T\int_0^{\infty}df\, 
\left( \frac{2\gamma(f)}{5}\right)^2
\frac{S_h^2(f)}{S_n^2(f)}
\right]^{1/4}\, .
\ee
The physical reason for the appearance 
of the factor 2/5 in the above formula will be clear
from eq.~(\ref{gamma}) below. 
(The number 2/5 is
specific to L-shaped interferometers, see below).
The  factor  of $2$ in front of the integral 
can instead be understood from 
$\int_{-\infty}^{\infty}df=2\int_0^{\infty}df$.

Finally, we mention
another useful formula which expresses $\hogw (f)$ in terms of the
number of gravitons  per cell of the phase space,
$n(\vec{x},\vec{k})$. For an isotropic stochastic background 
$n(\vec{x},\vec{k})=n_f$ depends only on the frequency
$f=|\vec{k}|/(2\pi)$, 
and $\rho_{\rm gw}=2\int n_f 2\pi f\, d^3k/(2\pi)^3 =
16\pi^2\int_0^{\infty}
d(\log f)n_ff^4$. Therefore  $d\rho_{\rm gw}/d\log f = 16\pi^2n_ff^4$, and
\be\label{37}
\hogw (f) \simeq 3.6\left( \frac{n_f}{10^{37}}\right)
\left(\frac{f}{\rm 1 kHz}\right)^4\, .
\ee
As we will discuss below, to be observable at the LIGO/Virgo
interferometers, we should have at least $\hogw\sim 10^{-6}$ between
1 Hz and 1 kHz, corresponding to $n_f$ of order
$10^{31}$ at 1 kHz and $n_f\sim 10^{43}$ at 1 Hz. A detectable
stochastic GW background is therefore exceedingly classical, $n_k\gg 1$.

\subsection{The characteristic  noise level}

We have seen in the previous section that
there is a very natural definition of the characteristic
amplitude of the signal, given by $h_c(f)$,
which contains all the informations on the physical effects, and is
independent of the apparatus. We can therefore associate to $h_c(f)$ 
a corresponding noise amplitude $h_n(f)$, that embodies
 all the informations
on the apparatus, defining  $\hc /h_n(f)$ in terms of the optimal SNR.

If, in the integral giving the optimal SNR, eq.~(\ref{SNR})
or eq.~(\ref{SNR2}), we consider
only a range of frequencies $\Delta f$ such that the integrand is
approximately constant, we can write
\be\label{SNR4}
{\rm SNR}\simeq  
\left[ \frac{8}{25} T\Delta f
\frac{\gamma^2(f)S_h^2(f)}{S_n^2(f)}\right]^{1/4}=
\left[\frac{2T\Delta f\gamma^2(f)h_c^4(f)}{25 f^2S_n^2(f)}\right]^{1/4}\, .
\ee
The right-hand side of eq.~(\ref{SNR4}) is proportional to $h_c(f)$,
and we can therefore define $h_n(f)$
equating the right-hand side of eq.~(\ref{SNR4}) to
$h_c(f)/h_n(f)$, so that 
\be\label{hn}
h_n(f)=\frac{1}{(\frac{1}{2}T\Delta f)^{1/4}}\,
\left[\frac{1}{2}fS_n(f)\right]^{1/2}
\left(\frac{5}{\gamma (f)}\right)^{1/2}\, .
\ee
For $L$-shaped interferometers, the
overlap function $\gamma (f)$ is defined in terms of the detector
pattern functions $F_{+,\times}^{(i)}$, where $i=1,2$ labels the
detector, as~\cite{Fla}
\be\label{gamma}
\gamma (f)\equiv\frac{5}{2}\int \frac{d\hat{\Omega}}{4\pi}\,
e^{2\pi i f\hat{\Omega}\cdot\Delta \vec{x}}
\left( F_+^{(1)}F_+^{(2)}+F_{\times}^{(1)}F_{\times}^{(2)}\right)\, .
\ee
If the separation $\Delta {x}$ between the two detectors is very
small, $2\pi f\Delta x\ll 1$
and if the detectors have the same orientation, so that
$F_{+,\times}^{(1)}=F_{+,\times}^{(2)}$
then (averaging also over the angle $\psi$ discussed in the previous
subsection)
\be\label{gam2}
\gamma (f)\simeq \frac{5}{2}\left(\langle F_+^2\rangle_{\hat{\Omega}}
+\langle F_{\times}^2\rangle_{\hat{\Omega}}\right) =
5\langle F_+^2\rangle_{\hat{\Omega}}\, .
\ee
For L-shaped interferometers $\langle
F_+^2\rangle_{\hat{\Omega}}=1/5$ (see ref.~\cite{Th}, eq.~(110),
or ref.~\cite{ST}). We 
see that the factor $5/2$ in the definition of $\gamma (f)$, 
eq.~(\ref{gamma}), has been inserted so that for parallel detectors at
the same site $\gamma (f)=1$. Therefore, 
the factor $5^{1/2}$ in eq.~(\ref{hn}) measures the increase in the
noise level due to the fact that stochastic GWs hit the detectors from
all directions, rather than just from the direction where the
sensitivity is optimal, while $1/\gamma^{1/2}$ measures the decrease in
sensitivity due to the detectors separation.

For a generic geometry, the factors 5 in the above formulas are
replaced by $1/\langle F_+^2\rangle_{\hat{\Omega}}$.
In the special case $\gamma (f)=1$ (detectors at the same site)
the factor $(5/\gamma )^{1/2}$ in eq.~(\ref{hn}) becomes therefore
$1/\langle F_+^2\rangle_{\hat{\Omega}}^{1/2}$, and we recover eq.~(66)
of ref.~\cite{Th} (for  comparison,  note that the quantity
denoted $S_h(f)$ in~\cite{Th} is called here $S_n(f)/2$).

From the derivation of eq.~(\ref{hn})
we can understand  the limitations implicit
in the use of  $h_n(f)$. It gives a measure of the noise level only
under the approximation that leads from eq.~(\ref{SNR2}), which is
exact (in the limit $h_i\ll n_i$),
to eq.~(\ref{SNR4}). This means that $\Delta f$ must be small enough 
compared to the scale on which the integrand in eq.~(\ref{SNR2}) changes,
so that $\gamma (f)S_h(f)/S_n(f)$ is approximately constant.
In a large bandwidth this is non trivial, and of course depends also on
the form of the signal; for instance, if $\hogw$ is flat,
then $S_h(f)\sim 1/f^3$.
For accurate estimates of the SNR at a wideband detector there is no
substitute for a numerical integration of eq.~(\ref{SNR})
or eq.~(\ref{SNR2}). However,
for  order of magnitude estimates, eq.~(\ref{rho5}) for $\hc$ and
eq.~(\ref{hn}) for $h_n(f)$ are simpler to use, and they have the
advantage of clearly separating the physical effect, which is
described by $\hc$, from the properties of the detectors, that enter
only in $h_n(f)$.

Eq.~(\ref{hn}) also shows  very clearly the advantage of correlating two
detectors compared with the use of a single detector. With a single
detector, the minimum  observable signal, at SNR=1,
is given by the condition
$S_h(f)\geq S_n(f)$. This means, from
eq.~(\ref{hc3}), a minimum detectable value for $h_c(f)$ given by
$h_{\rm min}^{\rm 1d}=(2fS_n(f))^{1/2}$. The superscript 1d reminds
that this quantity refers to a single detector. From eq.~(\ref{hn}) we
find for the minimum detectable value with two interferometers 
in coincidence, 
$h_{\rm min}^{\rm 2d}$, 
\be\label{gain}
h_{\rm min}^{\rm 2d}(f)=
\frac{1}{(\frac{1}{2}T\Delta f)^{1/4}}\,
\frac{1}{2}h_{\rm min}^{\rm 1d}(f)
\left(\frac{5}{\gamma (f)}\right)^{1/2}\simeq 
1.77\times 10^{-2}h_{\rm min}^{\rm 1d}(f)\,
\left(\frac{\rm 1\, Hz}{\Delta f}\right)^{1/4}
\left(\frac{\rm 1\, yr}{T}\right)^{1/4}\frac{1}{\gamma^{1/2}(f)}
\, .
\ee
Of course, the  reduction factor in the noise level is larger if
the integration time is larger, 
and if we increase the bandwidth $\Delta f$
over which a useful coincidence 
(i.e. $\gamma (f)\sim 1$) is possible.  Note that $\hogw$ is quadratic
in $h_c(f)$, so that an improvement in sensitivity by two orders of
magnitudes in $h_c$ means four orders of magnitude in $\hogw$.

\section{Application to various detectors}

{\bf Single detectors}. To better appreciate the importance of correlating
two detectors, it is instructive to consider first the sensitivity
that can be
obtained using only one detector. In this case a hypothetical signal
would manifest itself as an eccess noise, and should therefore satisfy
$S_h(f)\gsim S_n(f)$. Using eqs.~(\ref{hc3},\ref{ogwhc}) we write $S_h$
in terms of $\hogw$, and we write $S_n(f)=\tilde{h}_f^2$. For
the minimum detectable value of $\hogw$ we get
\be\label{single}
\hogw^{\rm min}(f)\simeq 10^{-2}\left(\frac{f}{\rm 100 Hz}\right)^3
\left(\frac{\tilde{h}_f}{10^{-22}{\rm Hz}^{-1/2}}
\right)^2\, .
\ee
Fig.~1, taken from
ref.~\cite{Gam},  shows the sensitivity of the planned Virgo
interferometer, i.e., the quantity $\tilde{h}_f$ as a function of $f$. 
We see that Virgo, used as a single detector, can
reach a minimum detectable value for $\hogw$
of order $10^{-2}$ or at most a few times
$10^{-3}$, at $f=$ 100Hz. 
Unfortunately, this is not an interesting
sensitivity level; as we will discuss in sect.~7, an interesting
sensitivity level for $\hogw$ is at least
of order $10^{-7}-10^{-6}$. To reach such a level with a single
Virgo interferometer we need, e.g., $\tilde{h}_f< 3\times 10^{-25}-
10^{-24} {\rm Hz}^{-1/2}$ at $f=$ 100 Hz, or $\tilde{h}_f<  10^{-26}-
3\times 10^{-26} {\rm Hz}^{-1/2}$ at $f=$ 1~kHz.
We see from fig.~1 that such small  values of $\tilde{h}_f$ are
very far from the sensitivity of  first generation interferometers,
and are in fact
even well below the  limitation due to quantum noise.

If we consider instead the NAUTILUS detector, which has a
 target sensitivity $\tilde{h}_f=8.6\times 10^{-23} {\rm Hz}^{-1/2}$
and operates at two frequencies in the kHz region (907 Hz
and 923.32 Hz)~\cite{Nau}, we find $\hogw^{\rm min}(f)\sim 5$.
The analysis  using existing EXPLORER
data (from 1991 and from 1994)  and more recent
NAUTILUS data  gives a bound $O(300)$ on $\hogw$~\cite{Nau2}.

Thus, we see that in these cases we 
cannot get a significant bound on $\hogw$. A very interesting
sensitivity, possibly even of order $\hogw\sim 10^{-13}$, could instead be
reached with a single detector, with the planned space interferometer
LISA~\cite{LISA}, at $f\sim 10^{-3}$ Hz.

Note also that, while correlating two detectors the SNR improves with
integration time, see eq.~(\ref{SNR}),
this is not so with a single detector. So, 
independently of the low sensitivity, with  a single detector it
is conceptually impossible to tell whether an eccess noise is due to a
physical signal or is a  noise of the apparatus that has not been
properly accounted for. This might not be a great problem if the SNR is
very large, but certainly with a single detector we cannot make a
reliable detection at SNR of order one, so that the above estimates
(which have been obtained setting SNR=1) are really overestimates. 

{\bf Virgo-Virgo}. 
We now consider the
sensitivity that could be obtained at Virgo if the planned
interferometer were correlated with a second identical interferometer
located at a few tens of kilometers from the first, and with the same
orientation.\footnote{Correlations between two interferometers have
already been carried out  using prototypes operated by the groups 
in Glasgow and at the
Max Planck Institute for Quantum Optics, with an effective coicident
observing period of 62 hours~\cite{Nic}. Although the  sensitivity  of
course is not yet significant, they demonstrate the possibility of
making long-term coincident observations with interferometers.} This
distance would be optimal from the point of view of the stochastic
background, since it should be sufficient to decorrelate local noises
like, e.g., the seismic noise and local electromagnetic disturbances,
but still the two interferometers would
be close enough so that the overlap function does not cut off the high
frequency range, as it happens, instead, correlating the two LIGOs.

\begin{figure}
\centering
\includegraphics[width=0.55\linewidth,angle=270]{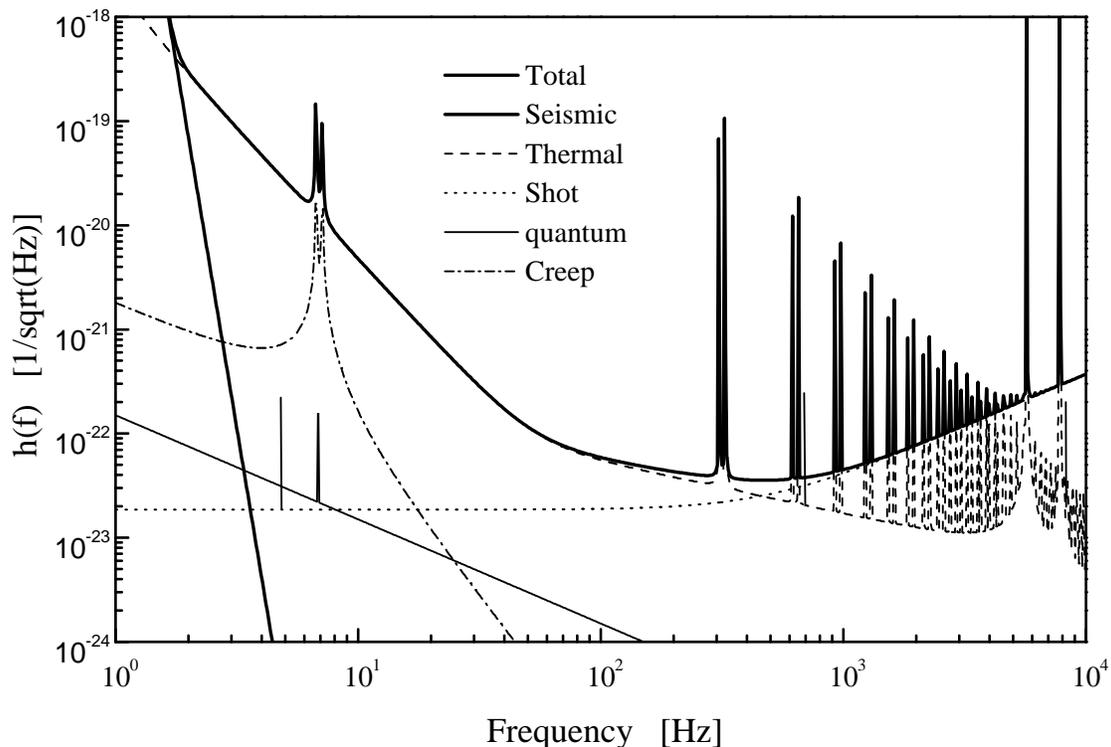}
\caption{The Virgo sensitivity curve (from ref.~[16]).}
\end{figure}

Let us first give a rough estimate of 
the sensitivity using $h_c(f),h_n(f)$.
From fig.~1 we see that we can take, for our estimate, 
$\tilde{h}_f\sim 10^{-22} {\rm Hz}^{-1/2}$ over a bandwidth
$\Delta f\sim$ 1 kHz. Using $T=$ 1~yr, eq.~(\ref{hn}) gives
\be
h_n(f)\sim 4.5\times 10^{-24}
\left(\frac{f}{100{\rm Hz}}\right)^{1/2}
\left(\frac{\tilde{h}_f}{10^{-22} {\rm Hz}^{-1/2}}\right)\, .
\ee
Requiring for instance  SNR=1.65 (this corresponds to $90\%$
confidence level;  a more precise discussion of the
statistical significance, including the effect of the false alarm rate
can be found in ref.~\cite{AR})
gives an estimate for the minimum detectable value
of $\hogw (f)$,
\be\label{min}
\hogw^{\rm min}(f)\sim 3\times 10^{-7}
\left(\frac{f}{100{\rm Hz}}\right)^{3}
\left(\frac{\tilde{h}_f}{10^{-22} {\rm Hz}^{-1/2}}\right)^2\, .
\ee
This suggests that correlating two Virgo interferometers we can detect
a relic spectrum with 
$\hogw ({\rm 100 Hz})\sim 3\times 10^{-7}$ at SNR=1.65, or
$1\times 10^{-7}$ at SNR=1.
Compared to the case of a single interferometer with SNR=1, 
eq.~(\ref{single}),
we gain five orders of magnitude.
As already
discussed, to obtain a precise numerical value one must however resort
to eq.~(\ref{SNR}). This involves an integral over all frequencies,
(that replaces the somewhat arbitrary choice of $\Delta f$ made above)
and depends on the functional form of $\hogw (f)$. If for instance
$\hogw (f)$ is independent of the frequency, using the numerical
values of $\tilde{h}_f$
plotted in fig.~1 (see ref.~\cite{Gam})
and performing the numerical integral we get for the minimum
detectable $\hogw$ (we give the result for a generic value of the SNR
and of the integration time)
\be\label{min2}
\hogw^{\rm min}\simeq 2 \times 10^{-7}\left(\frac{\rm
SNR}{1.65}\right)^2
\left(\frac{1\rm yr}{T}\right)^{1/2}
\hspace*{15mm} (\hogw (f)={\rm const.})\, .
\ee
We see that this number is quite consistent with the approximate
estimate~(\ref{min}), and with the value $2\times 10^{-7}$
reported in ref.~\cite{CS}. Stretching the parameters to
SNR=1 ($68\%$ c.l.) and $T=4$ years, the value goes down
at $(3-4)\times 10^{-8}$. 
It is interesting to note that the main contribution to the integral
comes from the region $f<$ 100 Hz. In fact, neglecting the
contribution to the integral of the region $f>$ 100 Hz, the result for
$\hogw^{\rm min}$ changes only by approximately $2\%$. Also, the lower
part of the accessible frequency range is not crucial. Restricting for
instance to the region 20 Hz $\leq f \leq$ 200 Hz, the sensitivity on
$\hogw$ degrades  by less than  $1\%$, while restricting
to the region 30 Hz $\leq f \leq$ 100 Hz, the sensitivity on
$\hogw$ degrades  by approximately $10\%$.
Then, from fig.~1 we conclude 
that by far the most important source of noise for the
measurement of a flat stochastic background is the 
thermal noise. In particular, the sensitivity to a stochastic background
is limited basically by the  mirror thermal
noise, which dominates in the region 40 Hz$\lsim f\lsim 200$ Hz, while
the  pendulum thermal noise  dominates below approximately 
40 Hz.\footnote{Note 
also that it is not very meaningful to give more
decimal figures in the minimum detectable
value of $\hogw^{\rm min}$. Apart from the
various uncertainties which enter the computation of the sensitivity
curve shown in fig.~1,
a trivial source of uncertainty is the fact that
the computation of the thermal noises are performed using 
a temperature of $300$K.
A $5\%$ variation, corresponding to an equally plausible 
value of the temperature,
gives a $5\%$ difference in $\hogw^{\rm min}$.
Quoting more figures is especially meaningless when the minimum
detectable $\hogw$ is estimated using the approximate quantities
$h_c(f),h_n(f)$, i.e. approximating the integrand of eq.~(\ref{SNR})
with a constant over a bandwidth $\Delta f$. For a broadband detector
these estimates typically 
give results which agree with the exact numerical integration 
of eq.~(\ref{SNR}) at best within a factor of two.}

The sensitivity depends however  on the functional form of 
$\ogw (f)$. Suppose for instance that
in the Virgo frequency band we can  approximate the signal as
\be
\ogw (f)=\ogw ({\rm 1kHz})\left(\frac{f}{\rm 1 kHz}\right)^{\alpha}
\, .
\ee
For $\alpha =1$ we find that
the spectrum is detectable at SNR=1.65 if $\hogw ({\rm 1kHz})\simeq
3.6\times 10^{-6}$. For $\alpha =-1$ we find (taking $f=$ 5Hz as
lower limit in the integration)
$\hogw ({\rm 1kHz})\simeq 6\times 10^{-9}$. Note however that in
this case, since $\alpha <0$, the spectrum is peaked at low
frequencies, and  $\hogw (5{\rm Hz})\simeq 1\times 10^{-6}$. So,
both for increasing or decreasing spectra, to be detectable
$\hogw$ must have a peak value, within the Virgo band, of order
a few $\times 10^{-6}$ 
in the case $\alpha =\pm 1$, while a constant spectrum can
be detected at the level $2\times 10^{-7}$.
Clearly, for detecting 
increasing (decreasing) spectra,  the upper (lower) part
of the frequency band becomes more important, and this is the reason
why the sensitivity  degrades  compared to flat spectra, since 
for increasing or decreasing spectra the maximum of the signal is at
the edges of the accessible frequency band, where the interferometer
sensitivity is worse.

\vspace{2mm}

{\bf LIGO-LIGO}. The two LIGO detectors are under construction at a
large distance from each other, $\sim 3000$ km. This choice optimizes
the possibility of detecting the direction of arrival of GWs from
astrophysical sources, but it is not optimal from the point
of view of the stochastic background, since the overlap functions cuts
off the integrand in eq.~(\ref{SNR}) around 60 Hz.
The sensitivity to a stochastic background  of the LIGO-LIGO detectors
has been computed in refs.~\cite{Mic,Chr,Fla,All,AR}. The result, for 
$\ogw (f)$ independent of $f$, is
\be
\hogw\simeq (5-6)\times 10^{-6}
\ee
for the initial LIGO. The advanced LIGO aims at $5\times
10^{-11}$. These numbers are given at $90\%$ c.l. in~\cite{All}, and a 
detailed analysis of the statistical significance is given in~\cite{AR}.

\hspace*{2mm}

{\bf Resonant masses and Resonant mass-Interferometer}. 
Resonant mass detectors  includes bars like NAUTILUS, EXPLORER and
AURIGA (see e.g. refs.~\cite{Nau,Coc} for  reviews). 
Spherical~\cite{Coc2,Coc3} and truncated icosahedron
(TIGA)~\cite{TIGA} resonant masses are also being developed or
studied.  The correlation between two resonant bars
 and between a bar and an interferometer
has been considered  in  refs.~\cite{VCCO,CS,ALS,APP,AFPR}. 
The values quoted in ref.~\cite{VCCO} are as follows (using 1 year of
integration time and SNR=1;
the mimimum detectable value of $\hogw$ grows as $({\rm SNR})^2$, so the
minimum detectable value at SNR=1.65 is about a factor of 3 larger).
Correlating the AURIGA and NAUTILUS detectors, with the present
orientation, we can detect 
$\hogw\simeq 2\times 10^{-4}$. Reorienting the detectors
for optimal correlation, which is technically feasible, we can reach 
$\hogw\simeq 8\times 10^{-5}$. For AURIGA-VIRGO, with present
orientation, $\hogw\simeq 2\times 10^{-4}$, and for 
NAUTILUS-VIRGO, $\hogw\simeq 3.5\times 10^{-4}$.
A three detectors correlation
AURIGA-NAUTILUS-VIRGO, with present orientation, would reach 
$\hogw\simeq 1\times 10^{-4}$, and with optimal orientation 
$\hogw\simeq 6\times 10^{-5}$. Although the improvement in
sensitivity in a bar-bar-interferometer correlation is not large
compared to a bar-bar or bar-interferometer 
correlation,  a three detectors
correlation would be important in ruling out spurious
effects~\cite{VCCO}. Preliminary results on a NAUTILUS-EXPLORER
correlation, using 12 hours of data, have been reported
in~\cite{Nau3}, and give a bound $\hogw\sim 120$.

Using resonant optical techniques, it is possible
to improve the sensitivity of interferometers at special values of the
frequency, at the expense of their broad-band sensitivity. Since bars
have a narrow band anyway, narrow-banding the interferometer improves
the sensitivity of a bar-interferometer correlation by about one order
of magnitude~\cite{CS}.

While resonant bars have been taking data for years, spherical
detectors are at the moment still at the stage of theoretical studies
(although prototypes might be built in the near future), but could reach
extremely interesting sensitivities. In particular, two spheres with a
diameter of 3 meters, made of Al5056, and located at the same site,
could reach a sensitivity $\hogw\sim 4\times
10^{-7}$~\cite{VCCO}. This figure  improves using a more dense
material or increasing the sphere diameter, but it might be 
difficult to build a heavier sphere. Another very promising
possibility is given by hollow spheres~\cite{Coc3}. The theoretical
studies of ref.~\cite{Coc3} suggest a minimum detectable value
$\hogw\sim 10^{-9}$ at $f=200$ Hz.

\section{The energy scales probed by relic gravitational waves}

Let us consider  the standard Friedmann-Robertson-Walker (FRW)
cosmological model, consisting of a radiation-dominated (RD) phase
followed by the present matter-dominated (MD) phase, and let us call
$a(t)$ the FRW scale factor. The RD phase goes backward in time until
some new regime sets in. This could be an inflationary epoch, e.g. at
the grand unification scale, or 
 the RD phase could go back in time until Planckian energies are
reached and quantum gravity sets in, i.e., until $t\sim t_{\rm
Pl}\simeq 5\times 10^{-44}$~s. If the correct theory of quantum
gravity is provided by string theory, the characteristic mass scale is
the string mass which is somewhat smaller than the Planck mass
and is presumably in the $10^{17}-10^{18}$ GeV region, and the
corresponding characteristic time is therefore one or two orders of
magnitude larger than $\tpl$.
The transition between the RD and MD phases takes place at $t=t_{\rm
eq}$, when  the temperature
of the Universe is of the order of only a few eV, so we are interested in
graviton production which takes place well within the RD phase, or
possibly at Planckian energies. 

A graviton produced with a frequency $f_*$ 
at a time $t=t_*$, within the RD phase has
today ($t=t_0$) a red-shifted frequency $f_0$ given by
$f_0=f_* a(t_*)/a(t_0)$. To compute the ratio $a(t_*)/a(t_0)$ one uses
the fact that during the standard RD and MD phases the Universe
expands adiabatically. The entropy per unit comoving volume is 
$S={\rm const.} g_S(T)a^3(t)T^3$, where $g_S(T)$ counts the effective
number of species~\cite{KT}. In the standard model, at
$T\gsim 300$ GeV, $g_S(T)$ becomes constant and has the value
$g_S(T_*)=106.75$,
while today $g_S(T_0)\simeq 3.91$~\cite{KT} and
$T_0=2.728\pm 0.002 $K~\cite{COBE}. Using
$ g_S(T_*)a^3(t_*)T_*^3= g_S(T_0)a^3(t_0)T_0^3$
one finds~\cite{KKT}
\be\label{f0}
f_0=f_*\frac{a(t_*)}{a(t_0)}\simeq  8.0\times 10^{-14}f_*
\left(\frac{100}{g_S(T_*)}\right)^{1/3}
\left(\frac{1{\rm GeV}}{T_*}\right)\, .
\ee
The first point to be addressed is what is the characteristic value of
the frequency $f_*$ produced at time $t_*$, when the temperature was
$T_*$.  One of the relevant
parameters in this estimate 
is certainly  the Hubble parameter at time of production,
$H(t_*)\equiv H_*$. This comes 
from the fact that $H_*^{-1}$ is the size
of the horizon at time $t_*$. The horizon size, physically, is the
length scale beyond which causal microphysics cannot operate (see
e.g.~\cite{KT}, ch.~8.4), and
therefore, for causality reasons, we  expect that the characteristic
wavelength 
of gravitons or any other particles produced  at time $t_*$ will
be of order  $H_*^{-1}$ or smaller.\footnote{On 
a more technical side, the deeper reason has
really to do with the invariance of general relativity
under coordinate transformations, combined with the expansion over a
fixed, non-uniform,
background. Consider for instance a scalar field $\phi (x)$ and
expand it around a given classical configuration, $\phi (x)=
\phi_0(x)+\delta\phi (x)$. Under a general coordinate transformation
$x\ra x'$, by definition  a scalar field transforms as
$\phi (x)\ra\phi '(x')=
\phi (x)$. However, when we expand around a given background, we keep
its functional form fixed and therefore under $x\ra x'$,
$\phi_0(x)\ra\phi_0(x')$, which for a non-constant field configuration,
is different from $\phi_0(x)$. It follows that 
the perturbation $\delta\phi (x)$  is not a scalar under general
coordinate transformations, even if $\phi (x)$ was a scalar. 
The effect becomes important for the Fourier components of $\delta\phi
(x)$ with a wavelength comparable or greater than the variation scale
of the background $\phi_0(x)$. (We are discussing a scalar field for
notational simplicity, but of course the same holds for the metric
tensor $g_{\mu\nu}$). In a homogeneous FRW background the only
variation is temporal, and its timescale is given by the
$H^{-1}$. Therefore modes with wavelength greater than $H^{-1}$ are in
general plagued by gauge artefacts. This problem 
manifests itself, for instance, when computing density fluctuations in
the early Universe. In this case one finds spurious modes which can be
removed with an appropriate gauge choice, see e.g. ref.~\cite{KT},
sect.~9.3.6 or ref.~\cite{Muk}.} 

Therefore, we write $\lambda_*=\epsilon H_*^{-1}$. The above argument
suggests $\epsilon\leq 1$.
During RD, $H_*^2=(8\pi /3)G\rho_{\rm rad}$. The
contribution to $\rho$ from a single species of
relativistic  particle with $g_i$ internal states (helicity, color,
etc.) is
$\rho_{\rm rad}=g_i(\pi^2/30)T^4$ for a boson
an $\rho =(7/8)g_i (\pi^2/30)T^4$ for a fermion.
Taking into account
that the i-th  species has in general a temperature $T_i\neq T$ if
it already dropped out of equilibrium, 
we can  define a function
$g(T)$ from $\rho_{\rm rad}=(\pi^2/30)g(T)T^4$. Then~\cite{KT}
\be\label{g*}
g(T)=\sum_{i={\rm bosons}}g_i\left(\frac{T_i}{T}\right)^4+
\frac{7}{8}\sum_{i={\rm fermions}}g_i\left(\frac{T_i}{T}\right)^4
\ee
The sum runs over relativistic species. This holds if a
species is in thermal equilibrium at the temperature $T_i$. If instead
it does not have a thermal spectrum (which in general
is the case for gravitons) we can still use the above equation, where
for this species $T_i$ does not represent a temperature but is defined
(for bosons) 
from $\rho_i=g_i(\pi^2/30)T_i^4$, where $\rho_i$ is the energy density
of this species.
The quantity $g_S(T)$ used before for the entropy
is given by the same expression as $g(T)$, with
$(T_i/T)^4$ replaced by $(T_i/T)^3$. We see 
that both $g(T)$ and $g_S(T)$  give a measure
of the effective number of species. For most of the early history of the
Universe, $g(T)=g_S(T)$, and in the standard model
at $T\gsim $ 300 GeV they have the common value $g_*=106.75$,
while today $g(T_0)=3.36,g_S(T_0)=3.91$~\cite{KT}. Therefore
\be\label{H*}
H_*^2=\frac{8\pi^3 g_*T_*^4}{90\mpl^2}\, ,
\ee
and, using $f_*\equiv H_*/\epsilon$, 
eq.~(\ref{f0}) can be written as~\cite{KKT}
\be\label{f1}
f_0\simeq 1.65\times 10^{-7}\frac{1}{\epsilon}
\left(\frac{T_*}{1\rm GeV}\right)
\left(\frac{g_*}{100}\right)^{1/6}\,{\rm Hz}\, .
\ee
This simple equation allows to understand a number of important points
concerning the energy scales that can be probed in GW experiments. 
The simplest estimate of $f_*$ corresponds to taking $\epsilon =1$ in
eq.~(\ref{f1})~\cite{All}. In this case,
we would find that a graviton observed today
at the frequency $f_0=$ 1Hz was produced
when the Universe had a temperature $T_*\sim 6\times 10^6$ GeV. Using
the relation between time and temperature in the RD phase,
\be
t\simeq \frac{2.42}{g_*^{1/2}}
\left(\frac{\rm MeV}{T}\right)^2\, {\rm sec}\, ,
\ee
we find that this 
corresponds to a production time $t_*\sim 7\times 10^{-21}$ sec,
and at time of
production this graviton had an energy $E_*\sim $ 0.3 MeV. 
At $f_0=$ 100Hz we get $t_*\sim 7\times 
10^{-25}$ sec, $T_*\sim 6\times 10^8$ GeV and
$E_*=$ 3 GeV. These would be, therefore, the scales relevant to Virgo
and LIGO. For a frequency $f_0=10^{-4}$ Hz, relevant to LISA, we
would get instead $t_*=7\times 10^{-13}$ sec, $T_*\sim 600$ GeV.

However, the estimate  $\lambda_*\sim H_*^{-1}$,  or $\epsilon\sim 1$,
can sometimes be incorrect 
even as an order of magnitude estimate. In the Appendix
we  illustrate this point
with two specific examples, one in which the assumption 
$\lambda_{*}\sim H_*^{-1}$ turns out
to be basically correct, and one in
which it can miss by several orders of magnitudes. Both examples will
in general illustrate the fact that the argument does not involve only
kinematics, but also the dynamics of the production mechanism.

From eq.~(\ref{f1}) we 
see that the  temperatures of the early Universe
explored detecting today relic GWs
at a  frequency $f_0$ are smaller  by a factor approximately equal
(for constant $g_*$) 
to $\epsilon$, compared to the  estimate with $\epsilon =1$.
Equivalently, a signal produced at a given temperature $T_*$
could in principle show up today in the Virgo/LIGO
frequency band when a naive estimate with $\epsilon =1$ suggests
that it falls at lower frequencies.

There is however another effect, which
instead  gives hopes of exploring the Universe at much
{\em  higher} temperatures than
naively expected, using GW experiments. In fact, the characteristic
frequency that we have discussed is the value of the cutoff frequency
in the graviton spectrum. Above this frequency, the spectrum decreases
exponentially~\cite{BD}, 
and no signal can be detected. Below this frequency,
however, the form of the spectrum is not fixed by general arguments. 
Thermal spectra have a low frequency behaviour
$\ogw (f)\sim f^3$, as we read from eq.~(\ref{37}) inserting a
Bose-Einstein distribution for $n_f$, so that, at low $f$, $n_f\sim
1/f$. However, below the Planck scale  gravitons interact too weakly 
to thermalize, and there is
no a priori  reason for a $\sim f^3$ dependence. The gravitons will
retain the form of the spectrum that they had at time of production,
and this is a very model dependent feature. However, from a number of
explicit examples and general arguments
that we will discuss below, we learn that spectra
 flat or almost flat over a large range of frequencies seem
to be not at all unusual. 

This fact has potentially important
consequencies. It means that, {\em 
even if a spectrum of gravitons produced
during the Planck era has a cutoff at frequencies much larger
than the Virgo/LIGO frequency band, still we can hope to observe in
the 10Hz--1kHz region the low-frequency part of these spectra.}
In the next subsection we will therefore discuss what signals can be
expected from the Universe at extremely high (Planckian, string
or GUT)  temperatures.

\section{Toward the Planck era?}
The scale of quantum gravity is given by the Planck mass, 
related to Newton constant by $G=1/\mpl^2$. More
precisely, since in the gravitational action enters the combination
$8\pi G$, we expect that the relevant scale is  the reduced
Planck mass $M=\mpl /(8\pi )^{1/2}\simeq 2.44\times 10^{18}$ GeV.
Using eq.~(\ref{f1}) with $T_*=M$ and $\epsilon =1$ gives
\be\label{400}
f_0\sim 400\left(\frac{g_*}{100}\right)^{1/6}\, {\rm GHz}\, ,
\hspace{30mm}( T_*= \mpl /\sqrt{8\pi})\, .
\ee
The dependence on $g_*$ is rather weak because of the power 1/6 in
eq.~(\ref{f1}). For $g_*=1000$, $f_0$ increases by a factor $\sim
1.5$ relative to $g_*=100$.  
For $T_*=\mgut\sim 10^{16}$ GeV, and $g_*\simeq 220$ (likely
values for a supersymmetric unification),
\be
f_0\sim 2\,{\rm GHz}\, ,\hspace{30mm}(T_*= M_{\rm GUT})\, .
\ee
Using instead 
the typical scale of string theory, $M_S$,
the characteristic frequency is between these two values, since $M_S$
is expected to be approximately
in the range $10^{-2}\lsim M_S/\mpl\lsim 10^{-1}$~\cite{Kap}.
Actually, in the estimate of $f_0$, precise
numbers  depend on details of the criteria
used. For instance, in the context of string cosmology~\cite{GV,peak}
it might be more meaningful to perform the estimate
setting $H_*\sim M_S$  rather than $T_*\sim M_S$, since
the Hubble constant during the
large curvature regime is fixed by stringy corrections, whose scale is
given by $M_S$~\cite{GMV}. 
In this case, from eq.~(\ref{H*}), the estimate for
$T_*$ is $T_*\sim (90/8\pi^3g_*)^{1/4} (\mpl M_S)^{1/2}$ instead
of $T_*\sim M_S$. Thus,  the dependence of $f_0$
on $g_*$ is different
($\sim g_*^{-1/12}$ instead of the factor $g_*^{1/6}$ that appears in
eq.~(\ref{400})),  
and there appears also a dependence on $M_S/\mpl$~\cite{peak}. 
However, numerically the difference is not very significant, since if
say $g_*=100$, then $(90/8\pi^3g_*)^{1/4}\simeq 0.25$ while for
$g_*=1000$, $(90/8\pi^3g_*)^{1/4}\simeq 0.14$, and this number is
partially compensated by  $(\mpl /M_S)^{1/2}$, that
ranges between approximately 10 and 3; thus, for $f_0$
the overall number with
the estimate $T_*\sim M_S$ or with $H_*\sim M_S$ can differ, say, by
a factor of two, which is anyway beyond the accuracy of these order of
magnitude estimates.

The typical frequency region  for  GUT/string/Planck physics is
therefore between the GHz and a few hundreds GHz. This is very far
from the region accessible to interferometers, which arrive
at most  to 10 kHz, see fig.~1.

One might consider the possibility of building a detector in the GHz
region to search for a signal from the Planck era. GW detectors 
based on microwave superconducting cavities
that could operate at these frequencies have indeed been studied in
the literature~\cite{PPR,Cav} in the years 1978-1980.  However,
from eq.~(\ref{rho5}) we see that  a level $\hogw\sim 10^{-6}$
corresponds, at $f=1$  GHz,  to $h_c\sim 10^{-30}$, which is many
orders of magnitude below the sensitivities discussed
in~\cite{PPR}. Still, given the great interest that a detector in this
frequency range could have, it could be worthwhile to investigate them
further.\footnote{These cavities could also be constructed so
that they operate in the MHz region~\cite{Pic}, 
where the requirements on $h_c$
are less stringent, and the wavelength is long enough to perform a
correlation between two detectors. 
A relic background detected at these frequencies
would be extremely interesting since it would be  unambiguously of
cosmological origin, see sect.~7B.}

It is clear from the above estimates of the characteristic frequency, 
that a necessary condition for
observing  at Virgo or LIGO a signal from the Planckian era
is that we have a
relic gravitational spectrum which is not peaked around the cutoff
frequency, but rather is  almost flat from the GHz region, where it
has the cutoff, down to at least the kHz region, where interferometers
can operate.

A thermal spectrum behaves as $\hogw\sim f^3$ and the value in the kHz
region would be a factor $\sim 10^{-18}$ smaller than a
peak value at the GHz, and
therefore would be obviously unobservable. However, as already remarked,
gravitons do not thermalize and there is no a priori reason for a
$\sim f^3$ spectrum. On the contrary, there are good general reasons
and various explicit examples favoring almost flat spectra, at least
over some range of frequencies.

A simple and instructive example has been given by
Krauss~\cite{Kra}. Consider a
global phase transition in the early Universe, 
associated with some scalar field $\phi$
that, below a critical temperature, gets a vacuum expectation value
$\langle\phi\rangle$. For causality reasons,
in an expanding Universe the scalar field
cannot have a correlation length larger than the horizon size. So,
even if the configuration which minimizes the energy is a constant
field, the field will be constant only over a horizon distance.
During the RD phase the horizon expands, and the
field will relax to a spatially uniform configuration within
the new horizon distance. This relaxation process will in general
produce gravitational waves, since there is no reason
to expect  $\langle T_{\mu\nu}\rangle$ for the classical field
just  entering the horizon to be spherically symmetric.
A simple estimate of the spectrum produced makes use of the fact that the
characteristic scale of spatial or temporal variation of the field is
given by the inverse of the
Hubble constant, $H^{-1}$, at the value of time under consideration. Then,
the energy density of the field is $\rho\sim (\partial\phi )^2
\sim \langle\phi\rangle^2H^2$. The total energy 
in a Hubble volume is $\rho H^{-3}
\sim \langle\phi\rangle^2H^{-1}$ and
the corresponding quadrupole moment is  $Q\sim (\rho H^{-3}) H^{-2}\sim
\langle\phi\rangle^2H^{-3}$, times a constant smaller than one, which
measures the non-sphericity of the configuration, and which is not very
relevant for order of magnitude estimates of the frequency
dependence. 
The energy liberated in GWs in a horizon
time is then given by~\cite{Kra}
\be
\Delta E\sim H^{-1}\times {\rm Luminosity}\sim
H^{-1}G\left(\frac{d^3Q}{dt^3}\right)^2\sim 
G\langle\phi\rangle^4H^{-1}\, ,
\ee
and the energy density of GW produced is
\be\label{kr}
\frac{d\rho_{\rm gw}}{d\log f}
\sim \frac{\Delta E}{H^{-3}}\sim G\langle\phi\rangle^4H^{2}\, .
\ee
This is the energy density  at a frequency $f\sim H$,  at a
value of time $t=t(f)$ which is the time  when the mode with frequency
$f$ enters the horizon. So, $d\rho/d\log f$ in the above equation is
evaluated at different values of time for different
frequencies. However, in the RD phase, the energy density
scales like $\rho\sim
1/a^4(t)\sim t^{-2}$ and $H\sim t^{-1}$ so that 
$\rho$ scales as $H^2$. Therefore, eq.~(\ref{kr}) means that, if we
evaluate $d\rho/d\log f$ at the same value of time, for all
frequencies which at this value of time are inside the horizon, the
spectrum is flat, $\hogw (f)\sim$ const.
This is a rather nice example of how a flat spectrum can follow simply
from the requirement of causality and simple dimensional estimates.

The best known example of an (almost) flat spectrum is the
amplification of quantum fluctuations in the inflation-RD phase
transition~\cite{Gri,Rub,AbH,All2,Sah}. In this case, 
during the inflationary
phase the horizon size $H^{-1}$ is approximately
constant and perturbations exit
from the horizon, since the physical wavelength scales as $\sim
a(t)$. The spectrum of quantum fluctuations for gravitational waves 
during inflation is (see e.g.~\cite{KT}, pg.289) 
$\Delta h_{+,\times}(f)\sim H/\mpl$, where $H$ is
the Hubble constant at time of crossing. Different
wavelengths cross at different time, but since $H$ does not vary during
inflation, this is really independent of $f$. When 
the mode is outside the horizon, its amplitude stays constant,
again for causality reasons. During the RD phase the horizon size
grows faster than $a(t)$ and so finally the perturbation reenters the
horizon, and its energy density appears in the form of gravitons. Its
energy density as it crosses back is $d\rho/d\log f\sim \dot{h}^2\sim
f^2(\Delta h)^2$. As in the previous example, this is the value
of $d\rho/d\log f$ at frequency $f$, at the value of time
 when this wavelength crosses back
inside the horizon, so it is evaluated at different times for
different frequencies, and a dependence $\sim f^2$ means that the
spectrum, when $d\rho/d\log f$ is evaluated at the same time for all
frequencies, is flat.
So, again in this example a basic ingredient for a flat spectrum
was causality, togheter with the fact that $H$ is constant during
inflation. 

The same mechanism of amplification of vacuum fluctuations produces an
interesting stochastic background in string cosmology.
In recent years
Gasperini and Veneziano have proposed an inflationary model based on
string theory~\cite{GV}. In this model the Universe starts at low
curvatures; at this stage it is described by the lowest order
effective action of string theory. Then it
evolves toward a regime of large curvature, where higher order
corrections to the string effective action stabilize the growth of the
curvature~\cite{GMV,MM}, which is now of order one in string units,
and the evolution
should then be matched to the standard RD phase~\cite{gra}.
The big-bang, 
i.e. a state of the Universe with Planckian energies and
curvature, rather than being a postulated initial condition, is the
end result of an evolution (the `pre-big-bang' phase)
which started from a cold Universe at low curvature.

The form of the spectrum of relic GWs has been discussed in
various recent papers~\cite{BGGV,bru,BMU,AB,Mau,revMau,peak}
and depends on the rate of growth of
the curvature and of the dilaton field in the various phases of the
model.
Very low frequencies are sensitive to the behavior in the low
curvature pre-big-bang regime. In this regime both the curvature and
the dilaton are growing and the spectrum turns out to grow as
$\hogw\sim f^3\log^2f$. In the large curvature regime, according to a
possible scenario~\cite{GMV} the Hubble parameter is constant, while
the dilaton can grow, with a derivative parametrized by an unknown
constant. In the limit in which the derivative of the dilaton in this
phase goes to zero, the spectrum in the region $f_s<f<f_{\rm peak}$ is
approximately flat. The value of $f_s$ is determined by the duration
of this intermediate large curvature phase ($f_{\rm peak}/f_s$ is in
fact the total red-shift during this phase),
and it is not calculable
without a detailed knowledge of stringy phenomena. 
With a sufficiently long string phase, 
it is possible to have $f_s $ smaller than, say, 1kHz.
The cutoff value
$f_{\rm peak}$ is instead in the region of a few hundred GHz, as
already discussed. Therefore, if in the large curvature regime we have
a long inflationary phase with an almost constant dilaton, we have
again a spectrum which is practically flat over a very large range of
frequencies.\footnote{Note that, while in standard inflationary models
a long inflationary phase at large curvature would provide too much
GWs at the frequecies relevant for the COBE bound (see below), this is
not the case for string cosmology because of the behaviour
$\sim f^3\log^2f$ that sets in at low $f$.}

Another example of relic GW spectrum which is almost flat over a huge
range of frequencies is provided by GWs produced by the decay of
cosmic strings~\cite{cosmic}.  In this case the spectrum has a peak
around $f\sim 10^{-12}$ Hz, where $\hogw$ can reach
a few times $10^{-6}$,
and then it is almost flat
over a huge range of frequencies, from $f\sim 10^{-8}$ Hz to
the GHz region, where it has the cutoff (fixed by the
arguments previously discussed, considering that the relevant scale
for cosmic string is $\mgut$).  The basic reason for such a behavior
is the scaling property of the string network, which says that a
single lengthscale, the Hubble length, characterizes all properties of
the string network~\cite{cosmic,All}. The network of strings evolves
toward a self-similar configuration, with small loops being chooped
off very long strings, and the typical radiation emitted by a single
loop has a wavelength related to the length of the loop.

In conclusion, from these examples we learn that GW spectra almost
flat  over a large range of frequencies are not at all unusual in
cosmology, and often the approximate flatness of the spectrum is a
consequence of rather general principles, like causality, or the
existence of a single lenghtscale in the problem.
Therefore, there is some
hope that we need not go in the GHz region to explore Planck scale
physics. 

\section{The effects of inflation}
In spite of the fact that the standard
cosmological model based on a RD and a MD phase is not contradicted by
any experimental data, inflation (i.e., a
stage of exponential expansion, or more generally 
a period characterized  by the condition
$\ddot{a}(t)>0$ on the FRW scale factor $a(t)$)
has many compelling features, see e.g. 
refs.~\cite{KT,Lin,Tur} for reviews.
At present, there is no `standard model' of inflation. The mechanism
is very general, and can be implemented in different models. 
In the models which have been `traditionally'
studied, inflation is driven by the potential energy of a scalar field. 

Independently of the details,  these models predict an
almost flat gravitational wave spectrum extending from some
high-frequency cutoff down to extremely small frequencies, $f\sim
10^{-16}$ Hz, and rising as $1/f^2$ as we go at even lower
frequencies, such as $f\sim 10^{-18}$ Hz, that
corresponds to wavelengths of the order of the present
Hubble distance $H^{-1}(t_0)$. 
The value of $\hogw (f)$ for this spectrum depends on the Hubble
constant during inflation, $H_{\rm DS}$, which in turn is related to the 
energy scale $M_{\rm infl}$ 
at which inflation takes place, $H_{\rm DS}\sim M_{\rm infl}^2/\mpl$.
A strong signal in GWs at  wavelengths $\lambda\sim H^{-1}(t_0)$ would
produce a stochastic red-shift on the frequencies of the photons of
the 2.7K radiation, and a corresponding fluctuation $\delta T/T$
in their temperature. The measurements of $\delta T/T$
from COBE therefore put an upper bound on the intensity of the
relic GWs produced at 
these long wavelengths, that translates into an upper bound
on the Hubble constant during inflation and on the scale $M_{\rm
infl}$  where inflation occurs~\cite{Rub,KWh}. 
Using COBE four-year data (and taking properly into
account the relation between the amplitude at low frequency and the
tilt of the spectrum, which is not exactly flat) gives an upper
bound~\cite{KWh,Tur}
\be\label{Minfl}
M_{\rm infl}<3.4\times 10^{16} {\rm GeV}\, .
\ee
A crucial consequence of this bound would be that we cannot see today a 
cosmological signal from the Planck scale. Every particle production 
which took place at
Planckian times has been exponentially diluted by the subsequent
inflationary stage. In this case the
 maximum energy that can be explored by
cosmological observations is the reheating temperature after
inflation, which is equal to $M_{\rm infl}$ in the case of a perfectly
efficient reheating, and is in general lower.

However, the above conclusion is not unescapable. We have already
considered above
the inflationary model based on string theory proposed by
Gasperini and Veneziano~\cite{GV}.
In this model we have automatically an inflationary evolution during
the pre-big-bang phase, triggered by the kinetic energy of the dilaton
field.
The bound~(\ref{Minfl}) in this case does not apply
because the spectrum
of relic gravitational waves is not flat over the whole 
frequency range, from  COBE frequencies  $f\sim 10^{-16}$ up to the
cutoff frequency, as in standard inflation, but rather goes like $\sim
f^3\log^2f$ at sufficiently small $f$, thereby 
automatically avoiding the COBE bound by
many orders of magnitude~\cite{BGGV}. 

Refs.~\cite{GV,initial} discuss
whether this inflationary phase can solve the kinematical problems of
standard cosmology without fine-tuned initial conditions.
In this paper we do not want to enter a discussion on the positive and
negative aspects of any specific model. We rather take the model of
refs.~\cite{GV} as an indication that the bound~(\ref{Minfl}) is not
necessarily unescapable, and we discuss in subsections A and B
two different scenarios 
which depend on whether this bound is obeyed or not.  
The amplification of vacuum fluctuations is common to both scenarios
and is discussed in subsection C, where we also consider the
possibility of combining the pre-big-bang scenario with
potential-driven  inflation. 

\subsection{Cosmological signals from the reheating era}
If we do have an inflationary phase at a scale $M_{\rm infl}$ that
satisfies the bound~(\ref{Minfl}), there
is no gravitational analog of the 2.7K radiation. At the Planck scale
photons and gravitons can be produced simply by thermal collisions,
with probably similar production rates. However, the 
particles produced in this way at the
Planck (or string) scale  are exponentially diluted 
by the subsequent inflationary phase, and the photons that we see
today in the cosmic microwave background radiation (CMBR)
have been produced during the reheating era which terminated the
inflationary phase. The highest energy scale that can be probed by
cosmological observations in this case is given by the reheating
temperature (the amplification of vacuum fluctuations can provide an
exception to this statement, see subsect.~C).

Since the reheating temperature $T_{\rm rh}<M_{\rm
infl}\ll \mpl$, thermal collisions are by now
unable to produce a substantial amount of
gravitons. However, in this case
relic GWs can  be produced  through
some  non-equilibrium  phenomenon connected with
the reheating process. 
For a process that takes place at the reheating temperature $T_{\rm
rh}$ the estimate of the characteristic frequency  is given by 
eq.~(\ref{f1}) with $T_*=T_{\rm rh}$. In principle the reheating
temperature can be between a minimum value at the  
TeV scale (since this
is the last chance for baryogenesis, via the anomaly in the
electroweak theory) and a maximum value $T_{\rm rh}\sim M_{\rm infl}
\lsim 3 \times 10^{16}$ GeV. In supersymmetric theories, the gravitino
problem give a further constraint $T_{\rm rh}\lsim 10^9$
GeV~\cite{KT,KM}. More precisely, ref.~\cite{KM} gives a bound on
$T_{\rm rh}\sim 10^6-10^9$ GeV for a gravitino mass 100 GeV--1~TeV,
and a bound $10^{11}-10^{12}$ GeV independent of the gravitino mass.

So, phenomena occurring at reheating will
manifest themselves with cutoff frequencies between $10^{-4}$  and
$10^9$ Hz,  if we set $\epsilon =1$ in eq.~(\ref{f1}).
In typical non-supersymmetric models the reheating temperature is
large, say $10^{14}$ GeV, corresponding to $f_0\sim (1/\epsilon )
10^7$ Hz, and in this case interferometers or resonant masses
could only look for their low-frequency tails. If instead
the rehating temperature is $T_{\rm rh}\sim 10^9$ GeV, non-equilibrium
phenomena taking place during reheating would give a signal just in
the LIGO/Virgo frequency band.

A promising mechanism for GW generation during reheating is  bubbles
collision, in the case when inflation terminates with a first order
phase transition~\cite{TuWi,KKT}. In this case the relevant estimate
is $f_0\sim (1/\epsilon )10^7$ Hz.
One should  however keep in mind
the possibility that bubble nucleation  occurs before the end of
inflation, and then the cutoff frequency would be red-shifted by
the subsequent inflationary evolution toward lower values, possibly
within the Virgo/LIGO frequency range~\cite{BAFO}. 

The other possibility is that  reheating occurs through the decay
of the inflaton field. In this case, in a very general
class of models, there is first an explosive stage called preheating,
where the inflaton field decays through a non-perturbative process
known as  parametric resonance~\cite{KLS}, and at this stage there are 
mechanisms that can produce GWs, see~\cite{param,Bas}. 
We believe that in these cases a reliable estimate of the
characteristic frequency is really very difficult to obtain, since the
relevant parameter, which is
the value of the Hubble constant at time of production, depends
on the complicated dynamics of preheating and on the specific
inflationary model considered. Ref.~\cite{param} examines models 
where the characteristic frequency turns out to be
in the region between a few tens and a few
hundreds  kHz.

\subsection{Cosmological signals from the Planck era}
The second possibility is that there is no
inflationary phase after the Planck era. This could happen if there is
no inflationary phase at all (although in this way we would lose the
advantages of inflation, so that this possibility does not
seem theoretically very
appealing), or if inflation is realized in the `pre-big-bang'
phase of string cosmology, as suggested in~\cite{GV}. Independently of
other considerations, the pre-big-bang model is interesting in this
context 
because it is an explicit example of the fact that physics at the
Planck or string scale can invalidate the considerations leading to
the bound~(\ref{Minfl}), without necessarily giving up the idea of
inflation.

If there is no inflationary phase after the Planck era, 
the photons of the CMBR have been produced at
the Planck (or string)
scale, when gravitons are also produced with probably similar rates,
and therefore we expect a GW analog of the 2.7~K radiation. At  Planckian
times both gravitons and photon will have approximately the same
characteristic energy. Then, gravitons decouple and their
characteristic energy scales like $1/a(t)$, while photons are in
thermal equilibrium and their temperature evolves so that the entropy
$\sim g_S(T)T^3a^3(t)$ is constant. Therefore the characteristic
energy of the gravitons today, $T_G$, is related to the 
present temperature of the photon, $T_0=2.728\pm 0.002$, by~\cite{KT}
\be
T_G=\left(\frac{3.91}{g_*}\right)^{1/3}T_0\simeq 0.93
\left(\frac{100}{g_*}\right)^{1/3} {\rm K}\, ,
\ee
where we used $g_S(T_0)\simeq 3.91$~\cite{KT} and $g_*$ is the 
effective number
of degrees of freedom at time of production. As we already discussed,
in the case of the gravitons $T_G$
should not be considered a temperature, since there is no 
a priori reason why
gravitons should have a thermal spectrum, but rather is a
characteristic energy, close to the high-energy cutoff in the spectrum.
The corresponding frequency is
$f_0\simeq 120\, (100/g_*)^{1/3}$ GHz; this is
the cutoff of a spectrum which extends  toward
lower frequencies. If its behaviour is, say, $\hogw\sim f^3$ there is
no hope of observing it in the kHz region, while if it is flatter it
can be more interesting for present experiments. 

Actually, since in this case we are
looking for a signal coming from the quantum gravity regime, the
situation can be richer. For instance, if the correct theory of
quantum gravity is provided by string theory, it may happen that
massive string modes are highly excited in the large curvature regime,
as indeed happens in the scenario discussed in ref.~\cite{MM}. In this
case the decay of excited string modes can be an interesting mechanism
for production of GWs. In fact, for large level number $N$ the decay
width of the string is dominated by transitions of the form $N\ra N-1$
with emission of a graviton or another massless particle~\cite{WTM}.
The mass levels of the string depend on $N$ as $M_N^2\sim N$ so that,
in string units, $\Delta M_N\sim 1/\sqrt{N}$, and therefore these
decays would produce a set of peaks  at today frequencies
$f_N=f_0/\sqrt{N}$. For large $N$ these frequencies are close to each
other and 
it is  possible that the width will be  large,
so that we will not have lines but rather a continuous spectrum,  but
in any case this mechanism shifts some of the energy from the GHz
region toward a more accessible frequency range.

Note also that 
the value $f_0\simeq 120 (100/g_*)^{1/3}$ GHz is really only
an upper bound, since the production through decay  of excited string
levels would take place before the onset of the RD phase, and a further
red-shift should be taken into account. This cannot be computed
in the absence of a detailed knowledge of the cosmological evolution
in the large curvature regime, but in general it might be possible
that the effect of the redshift and/or 
the factor $1/\sqrt{N}$ shift this
radiation in an interesting frequency band~\cite{MM}.

\subsection{Amplification of vacuum fluctuations}

A very general and well studied mechanism for the production of relic
GWs is the amplification of vacuum fluctuations due to a non-adiabatic
transition between different regimes for the evolution of the scale
factor. The mechanism is common to the scenarios discussed in the two
previous subsections.

A well known example is the transition between a (potential-driven)
inflationary phase and the standard RD phase~\cite{Gri,Rub,AbH,All2,Sah}. 
At first sight, it
is not promising for detection, since once the bound~(\ref{Minfl}) is
obeyed, at LIGO/Virgo frequencies
one gets  $\hogw\sim 10^{-15}$~\cite{Tur2}, very
far from the sensitivity of first and even of the planned second
generation  experiments. This is due to the fact that in this model
$\hogw$ is almost flat at all frequencies, from COBE frequencies 
up to the cutoff of the spectrum,
and therefore the bound imposed by COBE at
very low frequencies gives a bound at all higher frequencies.
One should however keep in mind that these computations are performed
assuming that before the transition the Universe is in the DeSitter
invariant vacuum. If instead before the transition
we have $n^0_k$ gravitons per cell of the
phase space,  the total number of gravitons after the transition is 
\be
n_k=n^0_k+2|\beta_k|^2\left( n^0_k+\frac{1}{2}\right)\, ,
\ee
where $\beta_k$ is the relevant 
Bogoliubov coefficient (we are considering the case
$\beta_{kk'}=\beta_k\delta_{kk'}$, as is usually the case in
cosmological  backgrounds).
If $n_k^0=0$ we have amplification of the `half
quantum' of vacuum fluctuations, with an amplification factor 
$2|\beta_k|^2$. But  also any preexisting non-vanishing $n_k^0$ is
amplified. One can expect that before the transition $n_k^0\ll 1$
because a phase space density of order one would correspond to a large
energy density, since at early time these modes have very high
frequency~\cite{AbH}. However, the
argument need not be true if there has been some mechanism which
pumped the energy into these modes, as may happen for instance 
with a previous amplification of vacuum fluctuations at the string
scale, as in~\cite{BGGV}.
For a transition between a DeSitter phase ($H$=const.)
and RD, the Bogoliubov coefficients depend on $k$ as 
$\beta_k\sim 1/k^2$~\cite{Gri,Rub,AbH,All2}; 
if before the transition $n_k^0=0$, then after
the transition $n_k=|\beta_k|^2\sim 1/k^4\sim 1/f^4$ 
and eq.~(\ref{37}) shows
that the spectrum for $\hogw (f)$ is flat (actually, during slow-roll
inflation, the Hubble parameter is not exactly constant, and this
results in a small tilt in the spectrum, see e.g.~\cite{SL}).  If
instead for some $k$ we have
$n_k^0\neq 0$ and $|\beta_k|\gg 1$, the dependence 
of $\hogw$ on $f$ is of the form $\hogw (f)\sim n_k^0+(1/2)$, since
the factor $f^4$ in eq.~(\ref{37}) is cancelled by $|\beta_k|^2\sim
1/f^4$. We see that {\em the spectrum of vacuum fluctuations after an
inflationary phase is determined by the occupation numbers before the
onset of inflation.} This is an example (actually, the only example
of which we are aware) of the fact that an
inflationary phase does not erase every information about the previous
stage of the Universe. A measurement of a relic GW spectrum, and hence
of $n_k$, if the Universe  underwent a stage of inflation, would allow
us to reconstruct the state of the Universe (i.e. $n_k^0$) {\em before}
inflation. 
The technical reason behind this is that the quantity which is
amplified is a number of particle per unit cell of the phase space,
and the volume of a cell of the phase space, $d^3xd^3k/(2\pi )^3$,
is unaffected by the expansion of the Universe,
contrarily to a physical spatial volume $d^3x\sim a^3(t)$.

These considerations can be applied to the case in which the pre-big-bang
model of Gasperini and Veneziano is followed, rather than by the
standard RD cosmology, by a phase of  inflation, driven by
the potential energy of some inflaton field. Indeed, it was stressed
already in the first papers on pre-big-bang cosmology~\cite{GV} that
the pre-big-bang scenario is by no means to be regarded as an
alternative to standard (even inflationary) cosmology, but rather it is a
completion of the standard picture beyond the Planck era. While most
efforts to date have been concentrated in matching the pre-big-bang
phase to a  RD phase, one can certainly consider the matching
to, say, a stage of chaotic inflation. This could have some attractive
features, since the inflationary stage that takes place during the
pre-big-bang phase, independently of whether it is long enough to
solve the horizon/flatness problems~\cite{initial}, can anyway prepare 
sufficiently homogeneous
patches that are  need to start chaotic inflation.
The spectrum of relic GWs in a model with a pre-big-bang
phase and a successive stage of potential-driven inflation would still
retain at low $f$ the form $\hogw\sim f^3\log^2 f$  obtained in string
cosmology, and therefore the comparison of the relic GW spectrum with
COBE anisotropies would give no bound on the scale $M_{\rm infl}$ 
of the  potential-driven inflation.

Another interesting example, considered in~\cite{GGV}, is the case
in which, before the onset of an inflationary
phase, the Universe was already in a state of thermal equilibrium, 
so that $n_k^0$ is given by a  Bose-Einstein distribution. 
This  is a natural initial condition  when the inflationary
phase is triggered by a first order phase transition, since this
requires a homogeneous thermal state as initial condition. 
This example works in the opposite direction compared to the 
case of string cosmology followed by
potential inflation. In fact, now
at small $k$ we have $n_k=(\exp (E_k/T)-1)^{-1}\sim 1/k$, 
and the spectrum is enhanced at low
frequency, and totally neglegible at Virgo/LIGO
frequencies. This also results in a  stronger bound on the
inflation scale compared to eq.~(\ref{Minfl})~\cite{GGV}.

A different possibility is that the equation of state in the early
Universe is modified, and this can result in growing
spectra so that again we can obey the COBE bound 
at $f\sim 10^{-16}$ Hz, but still we could have a large signal
at higher frequencies.
Ref.~\cite{Gri2} considers the case of a phase of the early
Universe with scale factor expanding as $a(\eta )\sim
|\eta|^{1+\beta}$, where $\eta$ is conformal time. This corresponds to
an equation of state $p=\gamma \rho$ with $\gamma =(1-\beta
)/(3+3\beta )$. Inflation corresponds to $\gamma =-1$ and then to
$\beta =-2$. At frequencies $f>10^{-16}$ Hz  the amplification of
vacuum fluctuations at the transition between this phase and the
standard RD phase gives a spectrum with $h_c(f)\sim f^{\beta +1}$
and therefore, from eq.~(\ref{ogwhc}), $\hogw\sim f^{2\beta +4}$,
which of course reproduces a flat spectrum in the inflationary case,
$\beta =-2$. The parameter $\beta$ is related to the spectral index of
density perturbations $n$ by $n=2\beta +5$, so that $\beta =-2$
corresponds to $n=1$, the  Harrison-Zeldovich spectrum. 
The value of $n$ can in principle be derived
from COBE observations. Two different groups, analyzing the same data
set in different ways, have produced the results $n=1.2\pm
0.3$~\cite{Ben} (therefore consistent with a Harrison-Zeldovich
spectrum) and $n=1.84\pm 0.29$~\cite{Bruk}. So, systematic errors
presumably dominate the analysis, but there is the possibility that
$n>1$ and therefore $\beta >-2$, which would give a growing
GW spectrum~\cite{Gri2}, $\hogw (f)  \sim f^{n-1}$. However, 
if $n$ is too large (like the value $n=1.84$) one
should also find a way to cutoff the spectrum before it grows too
large and exceeds existing bounds like the nucleosynthesis bound, see
sect.~7. 

If instead there is no phase of potential-driven inflation, while
there is an inflationary pre-big-bang phase as suggested in~\cite{GV},
than one finds  a relic GW spectrum that grows like $f^3\log^2f$ at
small $f$, thereby avoiding the COBE bound, and which has the cutoff
in the GHz region, as we already discussed. Whether the spectrum is
approximately flat from the GHz down to the kHz region relevant for
Virgo/LIGO depends on some
unknown parameters of the large curvature phase, in particular the
rate of growth of the dilaton and the duration of this phase. For some
values of these parameters (almost constant dilaton and long large
curvature phase) the spectrum is almost flat in a large
range of frequencies and it
might be relevant
at the frequencies of present
experiments~\cite{BGGV,bru,BMU,AB,Mau,revMau,peak}.

\section{Characteristic intensities}

\subsection{The nucleosynthesis bound}
Nucleosynthesis successfully predicts the primordial abundances of
deuterium, $^3$He, $^4$He and $^7$Li in terms of one
cosmological  parameter $\eta$,
 the baryon to photon ratio. In the prediction enter also
parameters of the underlying particle theory, which are therefore
constrained in order not to spoil the agreement. In particular, the
prediction is  sensitive to the effective number of species at
time of nucleosynthesis, $g_*=g(T\simeq {\rm MeV})$. 
With some simplifications, 
the dependence on $g_*$ can be understood as follows. 
A crucial parameter in the computations of nucleosynthesis is
the ratio of  the number
density of neutrons, $n_n$, to the number density  of protons, $n_p$, 
As long as thermal equilibrium is mantained we have (for
non-relativistic nucleons, as appropriate at $T\sim$ MeV, when
nucleosynthesis takes place)
$n_n/n_p=\exp (-Q/T)$ where $Q=m_n-m_p\simeq 1.3$ MeV.
Equilibrium is mantained by the process $pe\leftrightarrow n\nu$, with
width $\Gamma_{pe\ra n\nu}$, as long as $\Gamma_{pe\ra n\nu}>H$. When
the rate drops below the Hubble constant $H$, the process cannot
compete anymore
with the expansion of the Universe and, apart from
occasional weak processes, dominated by 
the decay of free neutrons, the ratio $n_n/n_p$
remains frozen at the value $\exp (-Q/T_{\rm f})$, where
$T_{\rm f}$ is the value of the temperature at time of freeze-out.  This
number therefore determines the density of neutrons available for
nucleosynthesis, and since practically all neutrons available will
eventually form $^4$He, the final primordial abundance of $^4$He is
very sensitive to the freeze-out temperature $T_{\rm f}$. Let us take for
simplicity $\Gamma_{pe\ra n\nu}\simeq G_F^2T^5$ (which is really
appropriate only in the limit $T\gg Q$). The Hubble constant
is given by  $H^2=(8\pi /3)G\rho$, where $\rho$ include all form of
energy density at time of nucleosynthesis, and therefore also the
contribution of primordial GWs. As usual, it is convenient to  write
the total energy density $\rho$ in terms of
$g_*$, see eqs.~(\ref{g*}), as $\rho =(\pi^2/30)g_*T^4$. 
We recall that, for
gravitons, the quantity $T_i$ entering eq.~(\ref{g*}) is defined by
 $\rho_{\rm gw}=2(\pi^2/30)T_i^4$, and this does not imply
 a thermal spectrum.
Then $T_{\rm f}$ is determined by the condition
\be
G_F^2T_{\rm f}^5\simeq \left(\frac{8\pi^3g_*}{90}\right)^{1/2}
\frac{T_{\rm f}^2}{\mpl}\, .
\ee
This shows that  $T_{\rm f}\sim g_*^{1/6}$, 
at least with the approximation that we used for 
$\Gamma_{pe\ra n\nu}$. A large energy density in
relic gravitons gives a large contribution to the total density $\rho$
and therefore to $g_*$. This results in a larger freeze-out
temperature, more available neutrons
and then in overproduction of $^4$He. This is the idea
behind the nucleosynthesis bound~\cite{Sch}. 
More precisely, since the density of $^4$He increases also with 
the baryon to photon ratio $\eta$, we could compensate an increase in
$g_*$ with a decrease in $\eta$, and therefore
we also need a lower limit on $\eta$, which is provided by the
comparison with the abundance of deuterium and $^3$He.

Rather than 
$g_*$, it is often used an `effective number of neutrino species'
$N_{\nu}$ defined as follows. In the standard model, at $T\sim $ 
a few MeV,
the active degrees of freedom are the photon, $e^{\pm}$, neutrinos and
antineutrinos, and they have the same temperature, $T_i=T$.  
Then,  for $N_{\nu}$ families of light neutrinos,
$g_*(N_{\nu})=2+(7/8)(4+2N_{\nu})$, where the factor
of 2 comes from the two elicity states of the photon, 4 from
$e^{\pm}$ in the two elicity states, and $2N_{\nu}$ counts the
$N_{\nu}$ neutrinos and the $N_{\nu}$ antineutrinos, each
with their single elicity state. For the Standard Model, $N_{\nu}=3$
and  therefore $g_*=43/4$. So
we can define an `effective number
of neutrino species' $N_{\nu}$ from
\be
g_*(N_{\nu})=\frac{43}{4}+
\sum_{i=\rm extra\, bosons}g_i\left(\frac{T_i}{T}\right)^4+
\frac{7}{8}\sum_{i=\rm extra\, fermions}g_i\left(\frac{T_i}{T}\right)^4\, .
\ee
One extra species of light neutrino, at the same temperature as
the photons, would contribute one unit to $N_{\nu}$, but all species,
weighted with their energy density, contribute to $N_{\nu}$, which of
course in general is not an integer.
For $i=$ gravitons, we have $g_i=2$ and
$(T_i/T)^4=\rho_{\rm gw}/\rho_{\gamma}$, where
$\rho_{\gamma}=2(\pi^2/30)T^4$ is the photon
energy density.
If gravitational waves 
give the only extra contribution to $N_{\nu}$, compared to the
standard model with $N_{\nu}=3$, using 
$g_*(N_{\nu})=2+(7/8)(4+2N_{\nu})$,
the above equation gives immediately
\be\label{NS}
\left(\frac{\rho_{\rm gw}}{\rho_{\gamma}}\right)_{\rm NS}
=\frac{7}{8}(N_{\nu}-3)\, ,
\ee
where the subscript NS reminds that this equality holds at time of
nucleosynthesis.
If more extra species, not included in the standard
model, contribute to $g_*(N_{\nu})$, then the equal sign in the above
equation is replaced by lower or equal.\footnote{To
compare with eq.~(56) of ref.~\cite{All}, note that 
while we use $\rho_{\gamma}$, in ref.~\cite{All} the bound is
written in terms of
the total energy density in radiation at time of nucleosynthesis,
which includes also the contribution of $e^{\pm}$, neutrinos and
antineutrinos,
$(\rho_{\rm rad})_{\rm NS}=[1+(7/8)(2+3)](\rho_{\gamma})_{\rm NS}$.}
The same happens if there is a contribution
from any other form of energy  present at the
time of nucleosynthesis and not included in the energy 
density of radiation, like, e.g., primordial black holes.

To obtain a bound on the energy density
at the present time, we note that from the time of
nucleosynthesis to the present time $\rho_{\rm gw}$ scaled as $1/a^4$,
while the photon temperature evolved following $g_S(T)T^3a^3=$
const. (i.e. constant entropy)
and $\rho_{\gamma}\sim T^4\sim 1/(a^4 g_S^{4/3})$. Therefore
\be
\left(\frac{\rho_{\rm gw}}{\rho_{\gamma}}\right)_0=
\left(\frac{\rho_{\rm gw}}{\rho_{\gamma}}\right)_{\rm NS}
\left( \frac{g_S(T_0)}{g_S({\rm 1\, MeV})}\right)^{4/3}=
\left(\frac{\rho_{\rm gw}}{\rho_{\gamma}}\right)_{\rm NS}
\left( \frac{3.913}{10.75}\right)^{4/3}\, ,
\ee
where the subscript zero denotes present time.
Therefore we get the nucleosynthesis bound at the
present time,
\be\label{227}
\left(\frac{\rho_{\rm gw}}{\rho_{\gamma}}\right)_0
\leq 0.227 (N_{\nu}-3)\, .
\ee
Of course this bound holds only for GWs that were  already produced at
time of nucleosynthesis ($T\sim$ MeV, $t\sim$ sec).
It does not apply to any stochastic
background produced later, like backgrounds of astrophysical origin.
Note that this is a bound on the total energy density in gravitational
waves, integrated over all frequencies. Writing $\rho_{\rm gw}=\int 
d(\log  f)\,
d\rho_{\rm gw}/d\log f$, multiplying both $\rho_{\rm gw}$ and
$\rho_{\gamma}$ in eq.~(\ref{227})
by $h_0^2/\rho_{c}$  and inserting the numerical value
$h_0^2\rho_{\gamma}/\rho_c\simeq 2.481\times 10^{-5}$~\cite{PDG}, we
get 
\be\label{nsbound}
\int_{f=0}^{f=\infty}d(\log f)\,\, \hogw (f)\leq
5.6\times 10^{-6}  (N_{\nu}-3)\, .
\ee
The bound on $N_{\nu}$ from nucleosynthesis is subject to various
systematic errors in the analysis, which have to do mainly with the
issues of how much of the observed
$^4$He abundance is of primordial origin,
and of the nuclear processing of $^3$He in stars, and as a consequence
over the last five years  have been quoted limits on $N_{\nu}$ ranging
from 3.04 to around 5. 
The situation has been recently reviewed in ref.~\cite{CST}. The
conclusions of ref.~\cite{CST} is that, until current astrophysical
uncertainties are clarified, $N_{\nu}<4$ is a conservative limit.
Using  extreme assumptions, a meaningful limit $N_{\nu}<5$ still
exists, showing the robustness of the argument.
Correspondingly, the right-hand side of eq.~(\ref{nsbound}) is,
conservatively, of order
$5\times 10^{-6}$ and anyway cannot exceed $10^{-5}$.

If the integral cannot exceed these values, also its positive definite
integrand $\hogw (f)$ cannot exceed it over an appreciable interval of
frequencies, $\Delta\log f\sim 1$. One might still have, in principle,
a very narrow peak in $\hogw (f)$ at some frequency $f$, with a peak
value larger than say $10^{-5}$, while still its contribution to the
integral could be small enough. But, apart from the fact that 
such a behaviour seems rather implausible, or at least is not
suggested by any
cosmological mechanism, it would also be probably of
little help in the detection at broadband detectors like Virgo, because 
even if we gain in the height of the signal we lose
because of the reduction of
the useful frequency band $\Delta f$, see eq.~(\ref{SNR}).

These numbers therefore give a first  idea of what can be considered an
interesting detection level for $\hogw (f)$, which should be at least a
few times $10^{-6}$, especially considering that the
bound~(\ref{nsbound}) refers not only to
gravitational waves, but to all possible sources
of energy which have not been included, 
like particles beyond the standard model, primordial black
holes, etc.

\subsection{The astrophysical background}
GWs of astrophysical origin are not subject to the nucleosynthesis
bound, and therefore our first concern is whether a stochastic
background due to a large number of unresolved astrophysical sources
can give a contribution to $\hogw (f)$ larger than  the
bound~(\ref{nsbound}), or anyway larger than the expected relic signal,
thereby masking the
background of cosmological origin. 

A first observation is that there is a maximum frequency at which 
astrophysical sources can radiate. This
comes from the fact that a source of mass $M$, even if very compact,
will be at least as large as its gravitational radius $2GM$, the bound
being saturated by black holes. Even if its
surface were rotating at the speed
of light, its rotation period would be at least $4\pi GM$, and the
source cannot emit waves with a period much shorter than that. 
Therefore we have a maximum frequency~\cite{Tho2},
\be
f\lsim \frac{1}{4\pi GM}\sim 
10^4\frac{M_{\odot}}{M}{\rm Hz}\, .
\ee
To emit near this maximum frequency an object must presumably
have a mass of the order of the
 Chandrasekhar limit $\sim 1.2 M_{\odot}$, which gives a maximum
frequency of order 10~kHz~\cite{Tho2}, and this limit can be saturated
 only by very compact  objects
 (see ref.~\cite{FP} for a recent review of GWs emitted in the
gravitational collapse to black holes, with typical frequencies 
$f\lsim 5$ kHz).
The same numbers, apart from factors of order one, can be obtained
using the fact that for a self-gravitating Newtonian system 
with density $\rho$, radius $R$ and mass $M=\rho (4/3)\pi R^3$,
there is a natural
dynamical frequency~\cite{Schu} 
\be
f_{\rm dyn}=\frac{1}{2\pi}(\pi G\rho )^{1/2}=
\left(\frac{3GM}{16\pi^2 R^3}\right)^{1/2}\, .
\ee
With $R\geq 2GM$ we recover the same order of magnitude estimate
apart from a factor $(3/8)^{1/2}\simeq 0.6$.
This is already an encouraging result, because it
shows that the natural frequency domains of cosmological and
astrophysical sources can be very different. We have seen that the
natural frequency scale for Planckian physics  is the GHz, while no
astrophysical objects can emit above,  say, 6-10 kHz.
A stochastic background detected above these frequencies would be
unambiguosly of cosmological origin.

However, ground based interferometers have their maximum sensitivity
around 100 Hz, where astrophysical sources hopefully produce
interesting radiation (since these sources 
were the original motivation for the construction of interferometers). 
The radiation from a single source is not a problem, since it is
easily distinguished from a stochastic background. The problem
arises if there are many unresolved sources. (Of course this is a
problem from the point of view of the cosmological background, but the
observation of the astrophysical
background would be very interesting in itself;
techniques for the detection of this background with a single
interferometer using the fact that it is not isotropic and exploiting
the sideral modulation of the signal have been
discussed in ref.~\cite{GBG}).

The stochastic background from rotating neutron stars has been
discussed in~\cite{Pos} and references therein. The main uncertainty 
comes from the estimate of the typical ellipticy $\epsilon$ of the
neutron star, which measures its deviation from sphericity. 
An upper bound on $\epsilon$ can be obtained assuming
that the observed slowing down of the period of known pulsars
  is interely due to
the emission of gravitational radiation. This is almost certainly a
gross overestimate, since most of the spin down is probably due to
electromagnetic losses, at least for Crab-like pulsars. 
With realistic estimates for $\epsilon$, ref.~\cite{Pos} gives, at
$f=100$ Hz, a value of $h_c(f)\sim 5\times 10^{-28}$, 
that, using eq.~(\ref{rho5}), corresponds to $\hogw (100 {\rm Hz})
\sim 10^{-15}$. This is very far from the sensitivity of even the
advanced experiments. An absolute upper bound can be obtained assuming
that the spin down is due only to gravitational losses, and this gives
$\hogw (100 {\rm Hz}) \sim 10^{-7}$, but again this value is probably a
gross overestimate. 

The stochastic background from supernovae is  studied in ref.~\cite{FMS}. 
The expected frequencies in this case are of the order of the kHz or
lower (down to 500-600 Hz, depending on the redshift when these
objects are produced). Using
the observational data on the star formation rate, it turns out that
the duty cycle, i.e. the ratio between the duration of a typical burst
and the typical time interval between successive bursts is low, of order
0.01. Therefore this background is not  stochastic, but rather
like a `pop noise', and can be distinguished from a really stochastic
background. The value of $\hogw$ for the background from supernovae 
have been computed in~\cite{FMS} assuming axially symmetric collapse,
and assuming that all sources have the same value of $a=J/(GM^2)$,
where $J$ is the angular momentum. The results depend on the value
of $a$, and on $h_0$. For typical choices, one gets values of order
$\hogw\sim 10^{-14}-10^{-11}$ at say $f=65$ Hz, rising up to 
$\hogw\sim 10^{-9}-10^{-8}$ around 3kHz, where the spectrum has the
maximum~\cite{FMS}. 

These results suggest that astrophysical backgrounds might not be a
problem for the detection of a relic background at LIGO/Virgo
frequencies. The situation is different in the LISA frequency 
band~\cite{Tho2,Schu,BenH,Pos,LISA}. LISA can reach a sensitivity of
order $\hogw\sim$ a few $\times 10^{-13}$ at $f\sim 10^{-3}$ Hz
(see fig.1.3 of ref.~\cite{LISA}). 
However, for frequencies below a few mHz, one expects a stochastic
background due to a large number of galactic
white dwarf binaries. The estimate
of this background depends on the rate of white dwarf mergers, which
is uncertain. With rates of order $4\times 10^{-3}$ per year (which 
should be a secure upper limit~\cite{Pos}), the background can be as
large as $\hogw\sim 10^{-8}$ at $f=10^{-3}$ Hz. This number is actually
quite uncertain, and in ref.~\cite{LISA}, fig.~1.3, it is used another
plausible rate, which gives for instance
$\hogw\sim 10^{-11}-10^{-10}$ at $f=10^{-3}$ Hz.
Above a 
frequency of order a few times $10^{-2}$ Hz,
most signals from galactic binaries can be resolved individually and
no continuous background of galactic origin is
presently known at the level of sensitivity of LISA.

It should be observed that, even if an astrophysical background is
present, and masks a relic background, not all hopes are lost. If we
understand well enough the astrophysical background, we can subtract
it, and  the relic background would still be observable if it is much larger
than the uncertainty that we have on the astrophysical background.  In
fact, LISA should be able to subtract the background due to white
dwarf binaries,
since there is a large number of binaries close enough to be
individually resolvable~\cite{LISA}. 
This should allow to predict with some
accuracy the space density of white dwarf binaries in other parts of
the Galaxy, and therefore to compute the stochastic background that
they produce. Furthermore, any background of galactic origin is likely
to be concentrated near the galactic plane, and this is another handle
for its identification and subtraction. The situation is more
uncertain for  the contribution of extragalactic  binaries, whcih
again can be relevant at LISA frequencies.
The uncertainty  in the merging rate is such that it cannot be
predicted reliably, but it is believed to be lower than the galactic
background~\cite{Pos}. In this case the only handle for the
subtraction  would be the form of the spectrum. In fact,
even if the strength is quite uncertain, the form of the spectrum may
be known quite well~\cite{LISA}. 

\subsection{Cosmological predictions for the intensity}
The next question is whether it is reasonable to expect that some
cosmological production mechanism saturates the nucleosynthesis bound. 
This of course depends on the production mechanism, but some general
considerations are possible. First of all, it is clear that, if GWs are
produced at the Planck scale by collisions and decays,
togheter with the photons that we observe today in the CMBR,
and there is not an inflationary phase at later time (the 
scenario of sect.~6B), we expect roughly $\rho_{\rm gw}\sim
\rho_{\gamma}$, so that the bound~(\ref{NS}) is approximately
saturated. 

If the mechanism  that produces GWs is different from the mechanism
that produces the photons in the CMBR, often it is still  possible 
to relate the respective energy densities, simply because
the scales of the two processes are  related.  
For instance, ref.~\cite{peak} discusses
the spectrum of relic GWs produced in
string cosmology, using only general arguments. The 
peak of the spectrum, $f_1$, is fixed with the
criterium $H_*\sim M_S$, which
as discussed in sect.~5, corresponds to an
effective  `temperature' at time of
production $T_*\sim (M_S\mpl )^{1/2}=(M_S/\mpl)^{1/2}\mpl$. 
Redshifting $T_*$ we get a characteristic
frequency today $f_1\sim (M_S/\mpl)^{1/2}T_0$. Fixing the 
peak energy from the `one graviton level'~\cite{peak}, $n_f\sim 1$,
eq.~(\ref{37}) gives a peak
energy density $d\rho/d\log f\sim f_1^4$, so that at the peak
frequency we have
\be\label{pe}
\frac{d\rho_{\rm gw}}{d\log f}\sim 
\left(\frac{M_S}{\mpl}\right)^2\\
\frac{d\rho_{\gamma}}{d\log f}\, ,
\ee
so that  $\rho_{\rm gw}$ is related to $\rho_{\gamma}$, although
with a suppression factor $(M_S/\mpl )^2$ which is expected to range
between $10^{-4}$ and $10^{-2}$, and numerical factors.

Similarly, in the production of GWs through bubble
collisions when inflation terminates with a first order phase
transition,  
there is only one energy  scale,  the vacuum energy  density  $M$
during inflation (so that $M^4$ is the false-vacuum energy density). 
The value of $M$ fixes the reheating temperature, and therefore
$\rho_{\gamma}$, from
$(\pi^2g_*/30)T_{\rm rh}^4\simeq M^4$, and 
fixes also the energy liberated in
gravitational waves. The latter can be estimated as follows~\cite{TuWi}.
The typical wavelength $\lambda_*$
produced in the collision of two bubbles will
be of the order of the radius of the bubbles when they collide, which
in turn is a fraction of the horizon scale $H_*^{-1}$.
As in sect.~4, we write  $\lambda_* =\epsilon H_*^{-1}$, and $H_*\sim
M^2/\mpl$. The energy liberated in GWs in the
collision of two bubbles is of order $E_{GW}\sim GM_B^2/\lambda_*$
where $M_B\sim M^4\lambda_*^3$ is the energy of a typical bubble. The
fraction of the false vacuum energy that
goes into GWs instead of going into $\rho_{\gamma}$
is therefore $E_{GW}/M_B\sim \epsilon^2$.  This gives, at time of
production 
\be\label{tw}
\rho_{\rm gw}\sim\epsilon^2\rho_{\gamma}\, .
\ee
Again $\rho_{\rm gw}$ is related to $\rho_{\gamma}$, and the basic
reason is that there is essentially only one dimensionful parameter
that enters in the estimate. 

From these explicit examples, we see that independently of the
production mechanism, eq.~(\ref{tw}) is quite general. The scale for
the intensity of a relic GW background is indeed fixed in many cases by
$\rho_{\gamma}$, which therefore gives a first order of magnitude
estimate of the effect. However, there are also suppression factors,
like the factor $\epsilon^2$ in eq.~(\ref{tw}) or
$(M_S/\mpl)^{2}$ in eq.~(\ref{pe}), that, togheter with the exact
numerical coefficients, are crucial for the detection at present
experiments. A value $\epsilon\sim 0.1$ would allow detection at the
level $\hogw\sim 10^{-7}$ while $\epsilon\sim 10^{-2}$ would require
$\hogw\sim 10^{-9}$, which is beyond the possibilities of first
generation experiments.

For GWs produced by collisions at the Planck era we can even expect
$\epsilon\sim 1$. However,
another type of suppression is present 
if we have a spectrum of relic GWs with a total energy density
$\rho_{\rm gw}\sim \rho_{\gamma}$ with a cutoff in the GHz, and we
want to observe it in the kHz region. In this case, as we discussed,
we must hope that the spectrum
is practically flat between the kHz and the GHz. While we have seen
in sect.~5
that there are good reasons for considering such a behaviour, the fact
that the energy density  is spread over a wide interval of
frequencies diminuishes of course the value of $\hogw (f)$ at a given
frequency. The nucleosynthesis bound~(\ref{nsbound}) 
then gives a maximum value at 1kHz 
\be\label{ns1}
\hogw (1 {\rm kHz})\leq \frac{5.6\times 10^{-6}(N_{\nu} -3)}
{\log\left(\frac{1 {\rm GHz}}{1 {\rm kHz}}\right)}\simeq 4\times 10^{-7}
(N_{\nu}-3)\, .
\ee
If the spectrum extends further toward lower frequencies, the maximum
value of $\hogw$ decreases accordingly, with a factor $\log (1{\rm
GHz}/ f_{\rm min})$ instead of $\log (1{\rm GHz}/ 1{\rm kHz})$.

In production mechanisms where
there is  no obvious relation between $\rho_{\rm gw}$ and
$\rho_{\gamma}$, there is no general reason suggesting a value of 
$\rho_{\rm gw}$  close to the nucleosynthesis bound, but
still we can have rather large relic backgrounds. An important example
is provided by cosmic strings
(see~\cite{cosmic} and references therein). Cosmic strings might form
during phase transitions, but whether they actually form or not
depends on the precise nature of the transition. Therefore, as
remarked in ref.~\cite{All},  the computation of the GW spectra from cosmic
strings must be considered as illustrative rather than
realistic (as on the other hand happens in most of the
examples that we have discussed).
Anyway, at LIGO/Virgo frequencies, the typical 
estimate of the intensity is of order $\hogw\sim 10^{-8}-10^{-7}$.
(Similar results are obtained from hybrid topological
defects~\cite{XMV}). The scale for these numbers is given first of all
by the combination $(G\mu )^2$, where $\mu$ is the mass per unit
length of the string. It comes from the fact that a loop radiates with
a power $P\sim \gamma G\mu^2$, ($\gamma$ is a dimensionless constant)
while, evaluating $\hogw$, another factor of $G$ comes from $1/\rho_c$.
For strings created at the GUT scale $G\mu\sim 10^{-6}$. Then, there
are also large 
numbers related to the number of strings per horizon volume,
which is thought to be of order 50, to the fact that 
also $\gamma\sim 50$,
and to the size of the loop at formation time~\cite{All}, that finally
give a value $\hogw\sim 10^{-8}-10^{-7}$. It is clear that in this
case the right order of magnitude can only be obtained with a detailed
understanding of the dynamics of the cosmic string network.

\section{Conclusions}
Present GW experiments have not been designed especially for the
detection of  GW backgrounds of cosmological origin. Nevertheless,
there are chances that in their frequency window there  might be a
cosmological signal. The most naive estimate
of the frequency range for signals from the
very early Universe singles out the
GHz region, very far from the region accessible to 
ground based interferometers,
$f<$ a few kHz, or to resonant masses. 
To have a signal in the accessible region, one of these two
conditons should be met: either we find a spectrum with
a long low-frequency tail, that extends from the GHz down to the kHz
region, or we have some explosive production mechanism much below the
Planck scale. As we have discussed, both situations seem to be not at
all unusual, at least in the examples that have been worked out to
date. The crucial point is
 the value of the intensity of the background.

With a very optimistic attitude, one could hope for a signal,
present just in the 10Hz-1~kHz band, with the maximum  intensity
compatible with the nucleosynthesis bound,  $\hogw\sim$ a few $\times
10^{-6}$ (or even $10^{-5}$, stretching all parameters to the maximum
limit). 
Such an option is not excluded, and the fact that such a
background is not predicted by the mechanisms that have been investigated
to date is probably  not a very strong
objection, given our theoretical ignorance of physics at the Planck scale
and the rate at which new production mechanisms have been
proposed in recent years, see the reference list. 
However, with more realistic estimates, on general grounds
it appears difficult to  predict  a
background that in the kHz region exceeds the level 
$\hogw\sim$ a few $\times 10^{-7}$, independently of the production
mechanism. This should be considered the minimum detection level at
which a significant search can start. Such a level is beyond the
sensitivity of first generation experiments, unless a ground based
interferometer is correlated with a second interferometer, located at
a distance small enough so that a significant correlation is possible,
and large enough to decorrelate local noises. A few tens of kilometers
would probably be the right order of magnitude. 
In this case we could detect a signal at the level $2\times 10^{-7}$
in one year of integration time, with SNR=1.65, and 
the level of a few $\times 10^{-8}$ could be reached
with longer
integration time (and possibly allowing for a 
slightly worse confidence level; this could make some sense in a stochastic
search because the SNR increases with time, and the hint of a signal at
low SNR would provide a strong motivation for pursuing the search).
The difficulties of such a detection are clear, but it should be
stressed that the payoff of a positive result would be enormous,
opening up a window in the Universe and in fundamental high-energy
physics that will never be reached with particle physics experiments.

\vspace*{5mm}

{\bf Acknowledgments.}
I am very grateful to Adalberto Giazotto for many interesting
discussions and stimulating questions, which prompted me to write down
this paper. I thank Valeria Ferrari and Raffaella Schneider for
discussing with me their unpublished results on astrophysical
backgrounds.  I also thank for useful discussions or comments on the
manuscript Pia Astone, 
Danilo Babusci, Carlo Baccigalupi, Alessandra Buonanno, 
Massimo Cerdonio, Eugenio Coccia, 
Stefano Foffa, Maurizio Gasperini, Marco Lombardi, Emilio Picasso,
Riccardo Sturani and  Andrea Vicer\`e.

\appendix

\section{}
In this Appendix we discuss the estimate of the quantity
$\epsilon$, that enters  eq.~(\ref{f1}), in two specific examples. The
discussion is meant to illustrate the uncertainty inherent in the
determination of $\epsilon$ and the range of values that it can take.

{\em i)} Inflation-RD transition.
Quantum particle creation due to vacuum fluctuations is
associated with a sudden change of the scale
factor, e.g. from a 
DeSitter inflationary phase to the RD phase~\cite{Gri,Rub,AbH,All2,Sah}. 
 Let us consider a typical inflationary
scenario~\cite{KT}: at some time $t_i$ the
potential energy $V(\phi )$ of some scalar field $\phi$ (the
`inflaton' field) starts to trigger inflation, and the Universe
expands exponentially. This inflationary stage ends at time $t_f$,
when the so-called slow-roll conditions on $V(\phi )$ are not anymore
satisfied. To solve the problems of  standard cosmology, $t_f-t_i$ should
be at least of order $60 H_{\rm DS}^{-1}$, where $H_{\rm DS}$ is the Hubble
constant during the DeSitter inflationary phase.
 At $t=t_f$ the temperature of the Universe is exponentially
small, and a reheating period must take place. In the simplest scenarios,
reheating lasts for a time 
interval $\Delta t\sim \Gamma_{\phi}^{-1}$, where $\Gamma_{\phi}$ is the
decay width of the inflaton, and the decay of the inflaton 
reheats the Universe, which then enters the standard RD phase.

The cutoff of the spectrum at large frequency 
is determined by how fast the transition between the two
phases takes place~\cite{All2,Sah}. Therefore 
the maximum value of the frequency produced is $f_*\sim 1/\Delta t
\sim \Gamma_{\phi}$. The gravitons have been produced  at a time
$t_*$, with $t_f\leq t_*\leq t_f+\Delta t$. If the transition proceeds
slowly, $\Delta t\gg t_f$, i.e. if $\Gamma_{\phi}$ is
sufficiently small compared to $\sim H_{\rm DS}$,
then $t_f+\Delta t\sim \Delta t$ and we can take $t_*\sim\Delta
t\sim\Gamma_{\phi}^{-1}$. Therefore
in this case the statement $f_*\sim H(t_*)=1/(2t_*)$ is indeed correct. 
However, if $\Gamma_{\phi}$ is sufficiently large, so that the
transition takes place on an interval $\Delta t\ll t_f$, the value of
time explored detecting these gravitons
is $t_*\simeq t_f$, while the characteristic frequency
produced is still $f_*\sim \Gamma_{\phi}$. In this case
$f_*\gg 1/t_*$, or $\lambda_*\ll H_*^{-1}$.

However, in most models $\Gamma_{\phi}\ll H_{\rm DS}$. For instance,
in models derived from supergravity,  a large $\Gamma_{\phi}$ would
produce a large reheating temperature $T_{\rm rh}$, and a value
$T_{\rm rh}>10^9$ GeV is forbidden because of the gravitino problem
(see e.g. ref.~\cite{KT}, pg. 298). And indeed small values of
$\Gamma_{\phi}$ come out quite naturally in supersymmetric models. For
instance, in the model of ref.~\cite{HRR},
$\Gamma_{\phi}\simeq\Delta^6/M^5$ where $M=\mpl/\sqrt{8\pi}$ and
$\Delta \simeq 3\times 10^{-5}M$ is a mass scale presumably related 
to supersymmetry breaking. Therefore in this case $\Gamma_{\phi}\sim
10^{-27}M$ is extremely small. 

Not all models of inflation satisfy $\Gamma_{\phi}\ll H_{\rm DS}$
automatically. For instance, in the old Coleman-Weinberg model
 $\Gamma_{\phi}> H_{\rm DS}$, but the model 
is not viable. So, in the case
of quantum production of particles after an inflationary phase,
the estimate $f_*\sim 1/t_*$, and therefore $\epsilon\sim 1$, 
is  correct, at least in typical viable
models.

{\em ii)} Phase transitions.
Another important example is given by
phase transitions in the early Universe, as the QCD
or the electroweak phase transition. These are believed to proceed
through bubbles nucleation of the low-temperature phase, when the
Universe cools below the critical temperature for the phase
transition, $T_c$.  If the transition occurs explosively, the
collisions of different  bubbles produce gravitational
waves~\cite{Hog,Wit}.  The characteristic wavelength of these GWs is of
the order  of the typical 
radius $R$ of the bubbles when they collide.
We write again  $\lambda_{\rm peak}(t_*)\sim R=\epsilon
H_*^{-1}$. Plausible values of $\epsilon$ have been discussed in
detail by Hogan~\cite{Hog} and by Witten (ref.~\cite{Wit},
app.~B). In the case of the QCD transition
$\epsilon\sim 1$ is excluded because otherwise
in the bubble  collisions there would be overproduction of primordial
black holes. Production of primordial black holes
 is severely constrained because the energy
density of primordial black holes scales like $1/a^3$ in the RD phase,
and therefore they would come to dominate the energy density of
the Universe. In particular, the Universe would not be radiation
dominated   during nucleosynthesis. Actually, primordial black holes
can evaporate through emission of Hawking radiation.  However, in
order to evaporate before nucleosynthesis, the scale at which
the phase transition takes place must be higher than $10^{11}$
GeV~\cite{TuWi}. Therefore $\epsilon$ could be of order one in phase
transitions which take place at these temperatures, but not in the QCD
or electroweak phase transition.

To estimate $\epsilon$ it is necessary to make
assumptions about how the phase transition is nucleated. If it is
nucleated by thermal fluctuations, ref.~\cite{Hog} suggests an upper
bound $\epsilon \lsim 10^{-2}$. For nucleation of bubbles via quantum
tunneling a detailed analysis has been done in ref.~\cite{TWW}. 
The relevant parameter is the nucleation rate, which 
again is a model-dependent quantity,
but for a wide class of models one typically finds $\epsilon\sim
10^{-3}-10^{-2}$~\cite{TWW,KoT,KKT}.  If the transition is nucleated
by impurities (which is most often the case, except in very pure and
homogeneous samples) the issue is much more complicated. For instance,
the impurities could be given by turbolent motion of the cosmic fluid
generated prior to the QCD epoch; they would depend on the detailed
spectrum of the fluid motion, or in general on the characteristic
distance between impurities, and in this case $\epsilon$ is basically
impossible to estimate, even as an order of magnitude.

\end{document}